\documentclass[12pt,usenatbib]{mn2e}
\usepackage{url,ulem,times,amsmath,amsfonts,amssymb,color,epsfig,epstopdf}
\usepackage{graphics}
\usepackage{subfigure} % input macros for figures
\usepackage{bm}
\usepackage{rotating}
\usepackage[T1]{fontenc}
\usepackage{ae,aecompl}
 \usepackage{array,multirow,graphicx}
\usepackage[percent]{overpic}
\usepackage{libertine}
\usepackage[libertine]{newtxmath}
\usepackage{pdfpages}

\usepackage{float}
\usepackage{hyperref}
\hypersetup{
   colorlinks=true,
   urlcolor=blue,
   citecolor=blue,
   pdfborder= 0 0 0}

\def\mpc{\,{\rm Mpc}}

\def\kpc{\,{\rm kpc}}

\def \indc{$\mathcal{C}$}

\def\eee{\,{\boldsymbol{e_3}}}

\def\away{\,{\textit{`away'}}}
\def\facing{\,{\textit{`facing'}}}
\def\ltrsim{\mathrel{\hbox{\rlap{\hbox{\lower4pt\hbox{$\sim$}}}\hbox{$<$}}}}
\def\gtrsim{\mathrel{\hbox{\rlap{\hbox{\lower4pt\hbox{$\sim$}}}\hbox{$>$}}}}

\newcommand{\myvect}{\protect\overrightarrow}

\begin{document}

\title[satellite shape alignment]
{The shape alignment of satellite galaxies in Local Group-like pairs from the SDSS}

%%%%%%%%%%%%%%
%         Authors
%%%%%%%%%%%%%%
\author[P. Wang et al.]
{Peng Wang$^{1,2}$ \thanks{E-mail: wangpeng@pmo.ac.cn},
Quan Guo$^{3}$,
Noam I. Libeskind$^{2,4}$ \thanks{E-mail: nlibeskind@aip.de},
Elmo Tempel$^{2,5}$,
\newauthor
Chengliang Wei$^{1}$,
Xi Kang$^{1}$ \thanks{E-mail: kangxi@pmo.ac.cn}\\
$^1$Purple Mountain Observatory (PMO), No. 8 Yuan Hua Road, Nanjing, 2100342, China \\
$^2$Leibniz-Institut f\"ur Astrophysik Potsdam (AIP), An der Sternwarte 16, 14482 Potsdam, Germany \\
$^3$Shanghai Astronomical Observatory (SHAO), Nandan Road 80, Shanghai 200030, China \\
$^4$University of Lyon, UCB Lyon 1/CNRS/IN2P3, IPN Lyon, France.\\
$^5$Tartu Observatory, University of Tartu, Observatooriumi 1, 61602 T\~oravere, Estonia}
%%%%%%%%%%%%%%%%%%%%%%%%%%%%%%%%%%%%%%%%%%%%%%%
%\date{Accepted --- . Received ---; in original form ---}
\pagerange{\pageref{firstpage}--\pageref{lastpage}} \pubyear{2018}
\maketitle
\label{firstpage}
%%%%%%%%%%%%%%%%%%%%%%%%%%%%%%%%%%%%%%%%%%%%%%%
%%%%%%%%%%%%%%
%          Abstract
%%%%%%%%%%%%%%
\begin{abstract}
It has been shown, both in simulations and observationally, that the tidal field of a large galaxy can torque its satellites such that the major axis of satellite galaxies points towards their hosts. This so-called `shape alignment'  has been observed in isolated Milky Way-like galaxies but not in `Local Group'-like pairs. In this study, we investigate the shape alignment of satellite galaxies in galaxy pairs similar to the Local Group identified in the Sloan Digital Sky Survey Data Release 13 (SDSS DR13). By stacking tens of thousands of satellite galaxies around primary galaxy pairs, we find two statistically strong alignment signals. (1) The major axes of satellite galaxies located in the (projected) area between two primaries (the {\it facing} region) tend to be perpendicular to the line connecting the satellite to its host (tangential alignment), while (2) the major axes of satellite galaxies located in regions away from the other host (the {\it away} region) tend to be aligned with the line connecting the satellite to its host (radial alignment). These alignments are confirmed at $\sim5\sigma$ levels. The alignment signal increases with increasing primary brightness, decreasing pair separation, and decreasing satellite distance. The alignment signal is also found to be stronger in filamentary environments. These findings will shed light on understanding the mechanisms of how satellite galaxies are affected by the tidal field in galaxy pairs and will be useful for investigating galaxy intrinsic alignment in the analyses of weak gravitational lensing.
\end{abstract}
%===========================================
%===========================================

%%%%%%%%%%%%%%
%       Introduction
%%%%%%%%%%%%%%
\section{Introduction}
\label{sec:intro}
According to the widely accepted $\Lambda$ cold dark matter ($\Lambda$CDM) model, the Universe was endowed with small primordial gravitational perturbations by a process known as inflation. Their power spectrum is constrained by the temperature inhomogeneities found in the CMB. These perturbations grow as the universe expands by gravitational instability \citep{1970A&A.....5...84Z} to form an inhomogenous distribution of matter on the largest scales. The density field of the universe constitutes a superposition of Gaussian random fields \citep{1986ApJ...304...15B} out of which CDM haloes collapse. Their growth occurs hierarchically; smaller haloes merge to form larger ones as the Universe evolves. When a smaller DM halo is subsumed by a larger one and becomes bound but not yet disrupted, it is termed a substructure or subhalo. DM haloes provide the potential in which gas may cool and form galaxies. As such galaxies merge with the mergers of their host haloes. Galaxies that are hosted by the main halo are known as central galaxies, while those galaxies that are hosted by subhaloes are called satellite galaxies.

Satellites are important probes to investigate the properties of their associated central galaxy and parent dark matter halo. For example, the kinematics of satellites can be used to constrain the dynamical masses of their host halo \citep{2002ApJ...571L..85M, 2003ApJ...593L...7B, 2003ApJ...598..260P, 2004MNRAS.352.1302V, 2007ApJ...654..153C}. In addition to as dynamical tracers, the spatial distribution of satellite galaxies also holds important information. For example, both observations \citep[e.g.,][]{1969ArA.....5..305H, 2006MNRAS.369.1293Y, 2007MNRAS.376L..43A, 2013ApJ...770L..12L}  and  numerical simulations \cite[e.g.,][]{2005ApJ...629L...5L, 2005ApJ...629..219Z, 2006ApJ...650..550A, 2010ApJ...709.1321A, 2005A&A...437..383K, 2007MNRAS.378.1531K, 2007MNRAS.374...16L, 2007ApJ...662L..71F, 2008ApJ...675..146F, 2009RAA.....9...41F, 2008MNRAS.390.1133B, 2009MNRAS.399..983A, 2011MNRAS.415.2607D, 2013SSRv..177..155L}
have focused on how satellites are distributed with respect to the shape of the central galaxies, in which they claimed that satellites (subhaloes) may have a statistical propensity to lie along the major axis of their central galaxy (host halo). However, this is not true in the local universe where the satellites around both the Milky Way and Centaurus A display polar orbits \citep{2018Sci...359..534M, 2005A&A...431..517K}. Moreover, satellite properties appear important when quantifying such alignments: \cite{2006MNRAS.369.1293Y} suggested a stronger alignment signal for red satellites of red central galaxies, while it is absent for blue satellites around blue central galaxies.

The spatial distribution of satellites described above is a natural consequence of the ellipsoidal shape of dark matter haloes, first found by \cite{2002ApJ...574..538J} n N-body simulations. The triaxial distribution of dark matter haloes is due to the anisotropic mass accretion on a large scale  \citep{1997MNRAS.290..411T, 2002ApJ...581..799V, 2004MNRAS.352..376A, 2004ApJ...603....7K, 2005MNRAS.364..424W, 2005ApJ...629..219Z}.

Quantifying the anisotropic nature of mass accretion has been the subject of a number of recent studies that have cast the problem in terms of examining the axes along which matter is compressed. Recent works suggest that more massive haloes accrete mass (mostly in the form of bound objects) {\it along}the axis of weakest collapse, while lower mass haloes accomplish most of their growth via diffuse, ambient, non-clumpy accretion that tends to occur {\it perpendicular} to the axis of weakest collapse \citep[see][]{2011MNRAS.411.1525L, 2014MNRAS.443.1274L, 2014ApJ...786....8W, 2015ApJ...807...37S,2015ApJ...813....6K, 2017MNRAS.468L.123W, 2018MNRAS.473.1562W}. Such a ``two phase'' accretion results in the well-established ``flipping'' of spin alignment \citep[see][and references therein]{2007ApJ...655L...5A, 2017MNRAS.468L.123W, 2018MNRAS.473.1562W, 2018MNRAS.481..414G, 2018ApJ...866..138W}. Consequently, satellites that survive accretion posses information regarding the geometry of the large scale structure.

One example of such an intimate relationship between satel- lites and the large-scale structure was shown by \cite{2015ApJ...800..112G} who found that galaxies in filaments have more satellites, and that these show a statistically significant alignment with the filament direction. This has been confirmed both using observation \cite{2017MNRAS.472.4769T} and simulation \citep{2015MNRAS.450.2727T}. \cite{2018ApJ...859..115W} first found strong environmental dependence of the spatial distribution of satellite around blue centrals.

In addition to the spatial distribution of satellites within a parent halo, a number of studies have focused on the ``radial alignment'' or  ``shape alignment'' of satellites: this is when the angle between the satellite (e.g. major) axis and its position relative to its host is examined [we refer the reader to section 3.4 of the thorough review of \cite{2015SSRv..193....1J} for more details). Observationally measuring the shape of satellites is complicated by a number of issues including unknown inclination angles, inherent triaxiality, and low luminosity. That said, \cite{1975AJ.....80..477H} and \cite{1976ApJ...209...22T} showed that in the Coma cluster the major axes of the member galaxies preferentially align with the direction towards the cluster centre, which suggested a radial alignment. During the early stages of the SDSS,  \cite{2005ApJ...627L..21P} and \cite{2008ApJ...675..146F} also found a radial alignment for satellites in cluster regions. However, subsequent studies \citep{2012MNRAS.421.3229H, 2013MNRAS.433.2727S, 2014MNRAS.445..726C, 2015A&A...575A..48S} have suggested distributions more consistent with random for shape alignments of satellites in groups and clusters. More recently,  \cite{2015MNRAS.450.2195S} selected satellites of bright early-type galaxies, and found that their semimajor axes tend to point towards their hosts.

By using subhaloes as a proxy for satellite galaxies, numerical simulations (unaffected by projection and inclination effects) often find a strong preferential radial alignment of satellite galaxies towards their host halo \citep{2007ApJ...671.1135K, 2008MNRAS.389.1266L, 2008ApJ...675..146F, 2008MNRAS.386L..52K, 2008MNRAS.388L..34K, 2008ApJ...672..825P, 2010MNRAS.405.1119K}. Among them, by using N-body simulations, \cite{2008ApJ...675..146F} claimed that the alignment is strongest within the virial radius of host halo and the signal drops off rapidly with increasing radius;\cite{2008MNRAS.389.1266L} suggested that the strength of the radial alignment increases with redshift and the mass of the host halo. A similar trend was also reported by \cite{2015MNRAS.448.3522T} in hydrodynamic cosmological simulations.

Similar to the origin of the satellite spatial distribution, the shape alignment of satellites is a dynamical effect \citep{2007ApJ...671.1135K, 2008ApJ...672..825P} caused by the tidal torque of the host halo \citep{1994MNRAS.270..390C}. A satellite galaxy can orbit inside a halo several times before it merges with its host galaxy. During this process, a satellite galaxy will be affected by the tidal field of the host’s dark matter halo, with a tendency to change its principal axis towards the host galaxies \citep[e.g.][]{2008MNRAS.388L..34K}. For a comprehensive picture of this process, we refer readers to fig. 8 in \cite{2008ApJ...672..825P}. \cite{2015MNRAS.447.1112B}  suggested that the major axis subhalo/satellite shows statistically a signal of radial alignment with the central galaxy.

As described above, {\it observational} investigations of the shape alignment have yet to reach a unified consensus. Regardless, the case of the satellite shape alignment in galaxy pairs still aroused our interest because our Local Group is dominated by a pair of primary galaxies: Milky Way and M31. A flattening of satellites distribution is found empirically in the Milky Way \citep[e.g.][]{1976MNRAS.174..695L, 2005A&A...431..517K, 2012MNRAS.423.1109P} and in M31 \citep{2013ApJ...766..120C, 2013Natur.493...62I, 2013MNRAS.435.1928P},  which is not an obvious prediction of generic galaxy formation models in the $\Lambda$CDM paradigm. The origin, stability and interpretation of such flattening satellite plane is still debated \citep[e.g.][]{2016MNRAS.462.3221A, 2015MNRAS.452.3838C, 2013MNRAS.431.3543H, 2010A&A...523A..32K, 2009MNRAS.399..550L, 2015MNRAS.452.1052L, 2014ApJ...790...74P, 2016ApJ...818...11S, 2014JGRG..119.2245Y, 2015MNRAS.452.1052L}. 

Galaxy pairs have been of interest to extragalactic astronomers since it was recognized nearly three decades ago that before galaxies merge, tidal interactions can trigger nuclear starbursts \citep{1987AJ.....93.1011K}, and enhance star formation rates by a factor of a few \citep[e.g.][]{2000ApJ...530..660B}. Indeed, a long series of over a dozen papers, focusing exclusively on how the properties of galaxies are affected by being paired up in the SDSS \citep{2008AJ....135.1877E, 2010MNRAS.407.1514E,2011MNRAS.418.2043E, 2013MNRAS.435.3627E, 2013MNRAS.430.3128E, 2015MNRAS.451L..35E, 2011MNRAS.412..591P, 2013MNRAS.433L..59P, 2016MNRAS.461.2589P, 2012MNRAS.426..549S, 2015MNRAS.449.3719S,  2014MNRAS.441.1297S, 2018MNRAS.476.2591V} has painted a detailed picture on the effect of the pair environment on star formation rates, colours, metallicity and active galactic nucleus (AGN) activity. A similar series of papers examining galaxy pairs in the 2dF survey has also accurately described the physics of galaxy pairs \citep{2003MNRAS.346.1189L, 2012A&A...539A..45L, 2006RMxAC..26..187A,2007MNRAS.375.1017A,2012A&A...539A..46A, 2008MNRAS.386L..82M}

The picture painted by these studies is that although many properties of paired galaxies are broadly consistent with the field, they do constitute a subclassification with statistically significant deviations from either field galaxies or larger group affiliations. For example, paired galaxies are more metal poor than they should be given their stellar mass \citep{2008AJ....135.1877E}; they have a younger stellar population \citep{2012A&A...539A..46A}; include more bluer galaxies than in the field \citep{2011MNRAS.412..591P,2012A&A...539A..45L}, and have a factor of 3 more AGN \citep{2011MNRAS.418.2043E}. We note that none of these studies have examined the differences in satellite population among paired galaxies. This is still an open question and is likely to remain so, until large samples of satellite galaxies can be identified with spectroscopy, as opposed to photometric methods (as in this work). That said, studies using photometrically identified satellites of galaxy pairs can begin to probe the effect of environment on satellite galaxies.

Therefore, the distribution of satellite in galaxy pairs is gaining increasing attention owing to the observational fact that the satellite galaxy system of M31 is highly lopsided, with about 80 per cent of its satellites lying on the side facing the Milky Way \citep{2013ApJ...766..120C}. Using SDSS data, \cite{2016ApJ...830..121L} found that there are up to $\sim10\%$ of satellites tend to bulge significantly towards the other central galaxy within the region between galaxy pairs. Similarly, \cite{2017ApJ...850..132P} have confirmed the lopsided distributed signal with Millennium simulations both in 2D and 3D space. \cite{2014ApJ...793L..42B} similarly showed that in general one would expect some degree of lopsidedness to possibly be observed, not from tides, but from the long memories of anisotropic group infall, although given the time-scales for dispersion it is unclear if such an argument can be applied to the samples seen in the SDSS. The flattening of satellite galaxies together with their lopsided distribution indicates that the actual shape of Milky Way halo may be significantly more complicated than the simple triaxial model we assumed. The shape alignment of satellites around galaxy pairs is another avenue along which such issues may be investigated. It can potentially provide information regarding the tidal field between galaxy pairs and how/if this is related to the stability and formation of thin satellite planes or lopsided distributions.

However, such an approach is challenging because the measurement of a satellite’s position angle is difficult owing to the inherent low surface brightnesses of these objects. Furthermore, any signal is likely limited by the fact that only a few satellites can be directly measured per primary galaxy. Owing to the advent of large galaxy surveys, for example, SDSS DR13 used in this work, a statistically robust treatment of satellite galaxies has become possible. In this paper, we investigate the shape alignment of satellites in galaxy pairs identified in the SDSS DR13, which are selected to roughly resemble our Local Group.

The outline of the paper is as follows. In Section 2, we describe the data and method including definitions of shape alignment and the estimation of uncertainties used in this work. In Section 3, we show the results of our analysis of the shape alignment between the direction of the projected major axis of satellites and its position vector with respect to their primary. In particular, we show how the shape alignment depends on the type of satellites, the magnitude of primaries, pair separation, and on the searching radius to find satellites. We also show the effect of large-scale filament on the shape alignment. Finally, we conclude and discuss our results in Section 4.

%%%%%%%%%%%
%       Method
%%%%%%%%%%%
\section{Data and Methods}
\label{sec:method}
In this section, we introduce how pairs of primary galaxies and their satellites are selected from the SDSS DR13 \citep{2000AJ....120.1579Y, 2017ApJS..233...25A} for our analysis, including how to select satellite galaxies based on the properties such as inverse concentration indices $\rm R_{90}/R_{50}$ and projected minor-to-major axial ratio $\rm b/a$.  Subsequently, we briefly introduce the filament catalogue used. Finally, we present the method used to quantify the alignment signal and the estimation of confidence intervals.

%------------------------------------------------------------------------------------------
\begin{figure*}
\centerline{\includegraphics[width=1.0\textwidth]{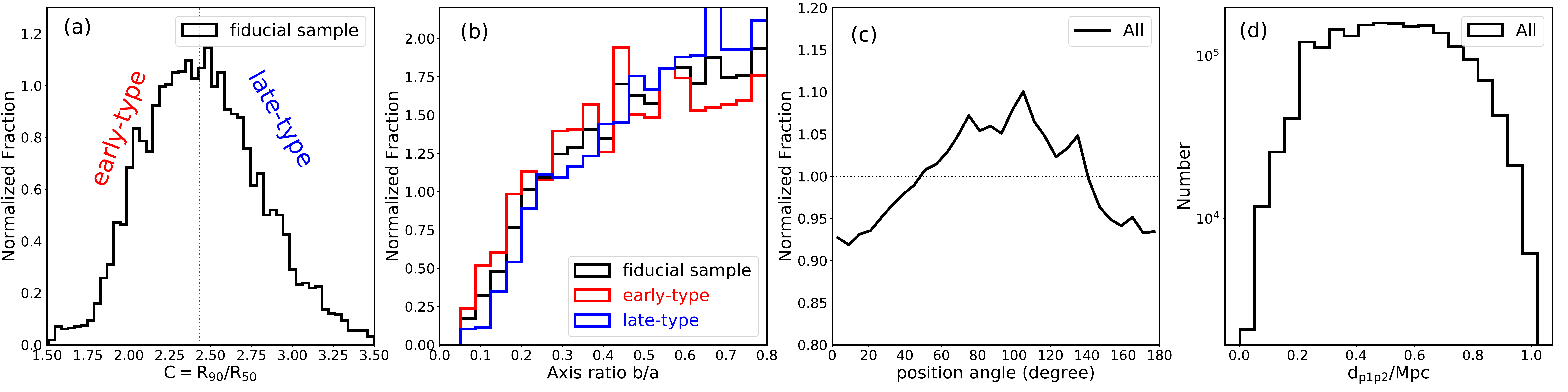}}
\caption{Panel-(a): the distribution of the normalized probability fraction of inverse concentration indices, $\rm \mathcal{C}=R_{90}/R_{50}$, which is used to determine the morphological type of satellite galaxies. The red dotted line indicates $\mathcal{C}=2.43$. Satellites with $\mathcal{C}<2.43$ are referred as `early-type' and otherwise `late-type'. Panel-(b): the  distribution of the axis ratio between the projected minor axis and the major axis, $\rm b/a$, for satellites. We only selected satellite with $\rm b/a<0.8$, above which galaxy position angle is too poorly defined to be included in this work. Panel-(c): the distribution of the normalized probability fraction of the position angle (the orientation of the projected major axis) of satellite galaxies. The dotted line indicates a uniform distribution. Panel-(d): the distribution of pair separation. Black solid line for all galaxy pairs we constructed from SDSS DR13, red solid line for the selected fiducial sample.}
\label{fig:method}
\end{figure*}
%------------------------------------------------------------------------------------------

\subsection{Galaxy pairs, satellites and filaments}
\label{sec:cata}

The sample of galaxy pairs is selected from the spectroscopic sample drawn from the SDSS DR13 with the same method used by \cite{2016ApJ...830..121L}. In this sample, the r-band magnitude of each member galaxy of galaxy pairs is limited at the range $[-21.5, -23.5]$, which is motivated by the magnitude range of the MW and M31 in the $r-$band \citep{2011ApJ...733...62L}. Note that different cuts of magnitude of galaxy pairs are examined in more detail in Section.~\ref{sec:mag}.  Meanwhile, the difference between the magnitudes of the two members of galaxies is also limited within 1 mag, and also with redshift differences of $\Delta_{z} < 0.003$.  The distance in the plane of the sky between the two primaries is required to be separated by $\rm 0.5<d_{p1p2}/Mpc<1.0$, which is also motivated by the distance between MW and M31. With such selection criteria, many of the pairs in our sample may not be physically bound as in previous studies (e.g., the 2DF series starting with \cite{2003MNRAS.346.1189L} or the SDSS series starting with \cite{2008AJ....135.1877E}). In general these studies are motivated by a desire to be very conservative in their pair selection, in order to ensure that they galaxy pairs are not contaminated by interlopers. Thus they tend to examine much smaller projected separations (e.g. less than $\sim$ 80 kpc/h in \cite{2008AJ....135.1877E} or less than 25kpc/h in \cite{2003MNRAS.346.1189L}). In our study, as we are interested in a lower limit of the shape alignment signal we are less concerned regarding the chances that our pair is not physical, seeing how any non-physical pairs will simply weaken any alignment signal we observe. Furthermore, we are motivated by the much larger separation in the Local Group  ($\sim750~kpc$). The effect of varying of the value of dp1p2 on the results is investigated in detail in Section.~\ref{sec:dis}. We selected 47 294 galaxy pairs for further cuts.

Potential smaller galaxies (hereafter refer as ``satellites'')  are searched from the full galaxy catalogue of the SDSS DR13. Although not strictly satellites in the bound sense, we refer to smaller galaxies that surround the hosts as ``satellites'' for simplicity, well aware that many of these may not in fact be bound to the pair. Our sample is a lower limit in the sense that we likely capture all the satellites of the pair (above our magnitude limit) among many interlopers that are not affiliated with the pair. This is done on purpose in order to get a lower limit for any shape signal. Since central galaxy pairs are selected from the spectroscopic galaxy sample, the potential satellites around the galaxy pairs are mostly fainter galaxies only with photometric redshift by construction. Following the method in \cite{2011MNRAS.417..370G, 2012MNRAS.427..428G}, we can search potential satellites from the photometric galaxy sample with a fainter magnitude limit ($m_r^{\rm lim}=20.5)$. We use the photometric redshift to help distinguish the interloper galaxies, which is similar to the methods used by \cite{2012MNRAS.421.1897O} or \cite{2014MNRAS.438.1784M}. Potential satellites candidates are chosen to be at least one magnitude fainter than the dimmer spectroscopically selected primary in the r-band and should be within the projected distance of  $R_{\rm search}$ and sufficiently close in redshift . `Sufficiently close'  is defined as a difference in spectroscopic redshift of less than $\Delta V = 600\ {\rm km\ s^{-1}}$ or, for galaxies without a spectroscopic redshift, with a photometric redshift within $\alpha_{\rm P}\sigma_{\rm P}$, where $\alpha_{\rm P }$ is a free parameter which is set to 2.5 according to the verification with the mock catalogues \citep{2011MNRAS.417..370G, 2013MNRAS.434.1838G} and $\sigma_{\rm P}$ is the photometric redshift error defined in the paper of \cite{2011MNRAS.417..370G}. We emphasize that a mere (less than $1\%$, see Table~\ref{table:number})  of our satellite galaxies have spectroscopic redshifts, in fact, confirmed satellites whose redshift is consistent with their hosts. Because this number is so low, there is no measurable effect on our results by eliminating them from our sample. Considering the credible position angle concentration indices of those potential satellites, we then kept nearly 1.9 million potential satellite galaxies. Each satellite is assigned to the closer primary galaxy of the pair according to the projected distance between them.

Note that we have also tested different samples of potential satellites where we  {\it ignore} the satellite photometric redshift and, purposefully, include interlopers during the searching. This implies that, assuming interlopers are randomly oriented, any signal we find will be diluted by the presence of interlopers and thus constitute a lower limit of any signal. We found that the resulting alignment signals are undetectable. For our fiducial cut, we select satellites with a searching radius ($R_{\rm search}$) of $100\kpc$, which is motivated by the MW satellites system wherein $64\%$ of known satellites are located within $100\kpc$ radius \citep[see][and reference within]{2015ApJ...813..109D}. We present the effect of different cuts in the searching radius in Section.~\ref{sec:searchR}.

Satellite galaxies are further split into early/late - type according to the inverse concentration indices $\mathcal{C} = \rm R_ {90} / R_ {50} $, where $R_{50}$ and $R_{90}$ are the Petrosian  half-light radius and  90\% light radius.  It is suggested that the concentration indices \indc~ correlate tightly with the morphology type of galaxies \citep{2001AJ....122.1238S}, while in this work we follow \cite{2011MNRAS.417..370G, 2012MNRAS.427..428G} and \cite{2011A&A...530A.106T} by using the inverse concentration indices \indc. The distribution of number faction of \indc~ is shown in the panel (a) of the  Fig.~\ref{fig:method}. For our analyses, we choose the value equals to 2.43 as the criterion to make sure that each type of satellites have the same number within $100 \kpc$. We refer satellite galaxies with $\mathcal{C} < 2.43$ as early-type and with $\mathcal{C}> 2.43$ as late-type. Noting that we also tested c=2.6, the value suggested in \cite{2011MNRAS.417..370G} as the criterion, we found that there is no significant effect on the results shown in Fig.~\ref{fig:type}.

In order to ensure the accuracy of the shape measurement of each satellite galaxy, we only consider satellite galaxy with $b/a\leq0.8$ (following the suggestion in \cite{2006MNRAS.369.1293Y} with $1-b/a\geq0.2$) since one would expect the orientation angle of these satellites to be the most accurate. Here $b/a$ is defined as the ratio between the minor and major axes of the 25-mag $\rm arcsec^{-2}$ isophote in the r band of the image of the satellite galaxy.  As shown in the panel (b) of Fig.~\ref{fig:method},the distribution of axial ratio of the early-/late-type satellite galaxies is consistent with all galaxies in the sample. The same distribution indicates that this cut introduces no systematic bias into either the early- or late-type samples.

%------------------------------------------------------------------------------------------
\begin{figure}
\centerline{\includegraphics[width=0.48\textwidth]{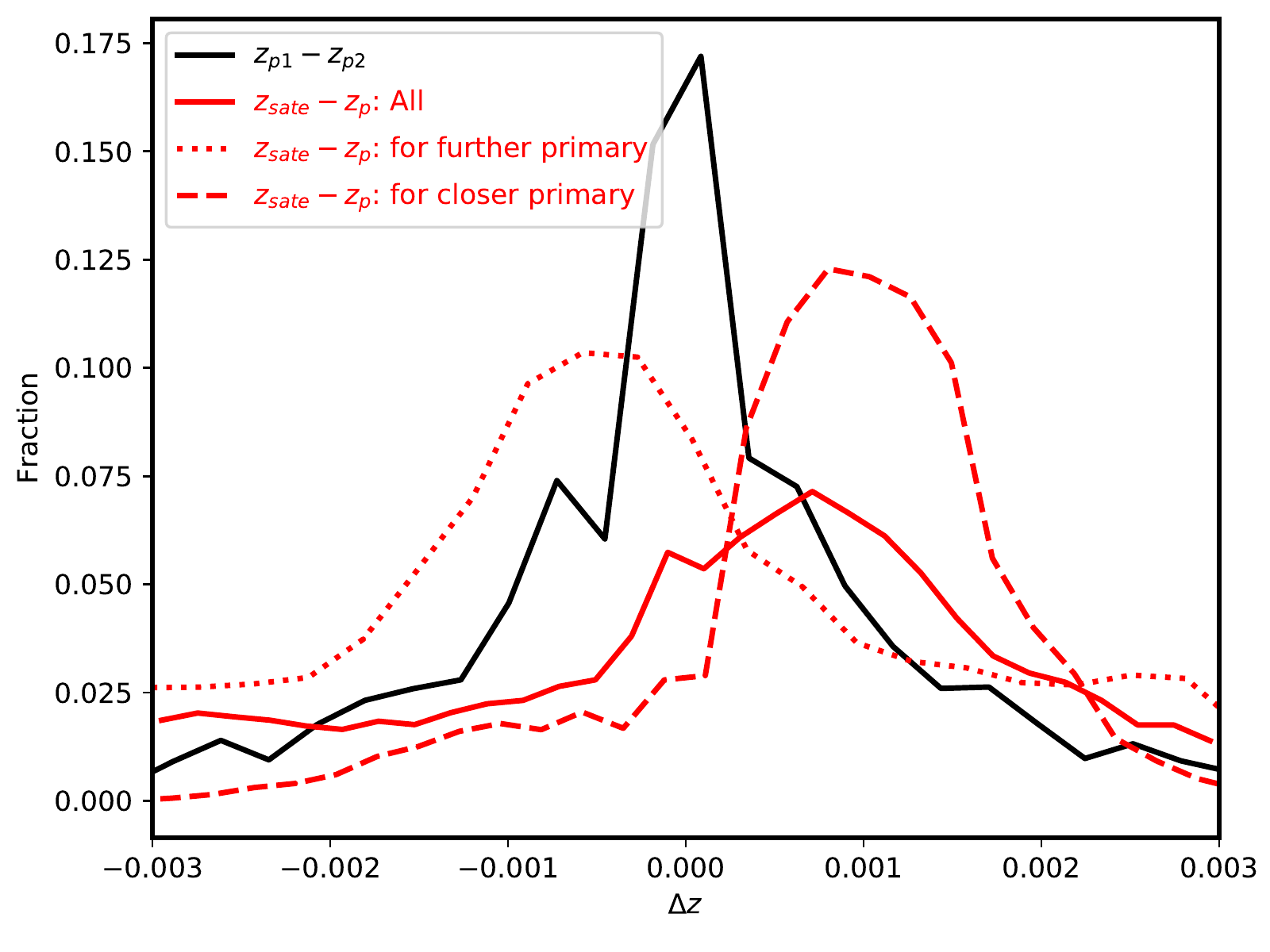}}
\caption{The distribution of redshift differences between members of pairs (black line), and between satellites and their primaries (red lines). Note that in order to make a robust result, those satellites with spectra redshift are used.}
\label{fig:diff_z}
\end{figure}
%------------------------------------------------------------------------------------------

%------------------------------------------------------------------------------------------
\begin{figure}
\centerline{\includegraphics[width=0.48\textwidth]{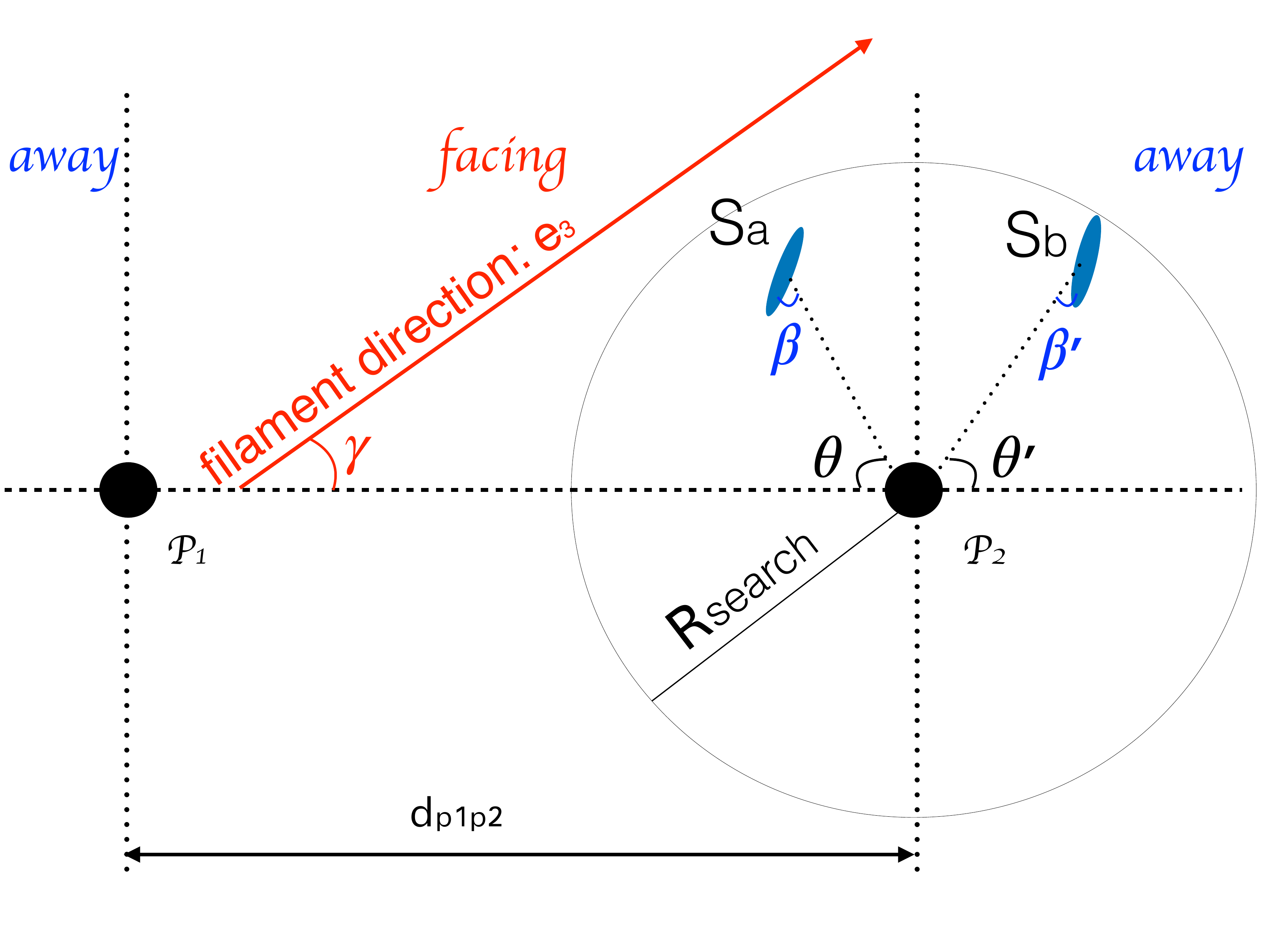}}
\caption{A schematic diagram showing how the system is projected in the plane of the sky. `$\rm P_{1}$' and `$\rm P_{2}$' are the two spectroscopically selected host galaxies, referred to as ``primaries'' (black points) with a projected separation distance $\rm d_{p1p2}$. `$\rm S_{a}$', `$\rm S_{b}$' are two satellites (blue ellipses, where the long axis represents the major axis) belong to this system within the searching radius, $\rm R_{search}$ and fulfilling the magnitude constraints so as not to find bright satellites. The line connecting  the two primaries is used to compute the angle, $\theta(\theta^{{\prime}})$, the distribution of which was the focus of \citet{2016ApJ...830..121L}. $\beta$ ($\beta^{\prime}$) is the angle between the orientation of the projected major axis of the satellite and its position vector relative to its primary. In this case, we show the system in filament environment and the angle between the  connecting line and filament direction $\eee$ (red arrow) is marked as $\gamma$. See text for more details about the definition of each angle.}
\label{fig:schematic}
\end{figure}
%------------------------------------------------------------------------------------------

The filament catalogue is built by applying an object point process with interactions (the Bisous process) to the distribution of galaxies in the spectroscopic galaxy sample \citep{2014MNRAS.437L..11T} from SDSS DR13, which is the same method used by \cite{2016ApJ...830..121L} as well. Random small segments (cylinders) based on the positions of galaxies are used to construct a filamentary network by examining the connectivity and alignment of these segments. A filamentary spine can then be extracted based on a detection probability and filament orientation \citep[see][also see \cite{2018MNRAS.473.1195L} for a comparison of Bisous model with other commonly used cosmic web classification methods]{2014MNRAS.437L..11T, 2016A&C....16...17T}. The radius of filaments in the catalog is assume to be roughly 0.7 Mpc, meaning that primary galaxies in filaments are less than 0.7 Mpc from the filament spine. This catalog is constructed with lower and upper CMB-corrected distance limits of z = 0.009 and z = 0.155, respectively. We thus confine our analysis to this redshift range. Filaments at higher redshift than the upper limit are too ``diluted'' to be detected. We classify a galaxy as ``in a filament'' if the distance of the galaxy from the axis of the filament is less than 0.7 Mpc and it is inside the filamentary cylinder.

To sum up, the criteria that determine the primaries and satellite galaxies are listed below:
\begin{itemize}
  \item host magnitude: $[-21.5, -23.5]$;
  \item magnitude difference: $\rm \Delta_{M_{12}}<1$ mag;
  \item host projected separation: $\rm 0.5<d_{p1p2}/Mpc<1.0$;
  \item searching radius: $\rm <100 \kpc$
  \item satellite axis ratio: $\rm b/a<0.8$.
  \item satellite type: early-type with $\rm \mathcal{C} < 2.43$, late-type with $\mathcal{C} > 2.43$;
  \item satellite magnitude: $<1$ mag fainter than the dimmer primary.
\end{itemize}   

Our final fiducial sample consists of 6268 pairs of primary galaxies with 5361 ($\rm < 100 \kpc$), 15700 ($\rm < 200 \kpc$) and 30099 ($\rm < 300 \ kpc$) potential satellites (also see Table~\ref{table:number}). The mean redshift difference of the pairs (namely, $\Delta_z = z_{P1} - z_{P2}$, where $z_{P1}$ and $z_{P2}$ are the redshifts of the brighter and fainter primary) is about $7.2\times10^{-4}$, which is similar to the sample used by \cite{2016ApJ...830..121L}. These pairs tend to be close (by construction) with redshift differences of $\Delta_z < 0.003$. The distribution of redshift differences between member of pairs is shown in black line in the Fig.~\ref{fig:diff_z}. The distribution is symmetric and around 88\% of galaxy pairs are within redshift difference $|z_{P1} - z_{P2}| < 0.0015$ ($\sim450 \ \rm km \ s^{-1}$).  The distribution of redshift differences between satellites and their primaries is shown in red line in the Fig.~\ref{fig:diff_z}. Note that only satellites with spectra redshift are used. It is interesting to note that this distribution is asymmetrically skewed towards positive $\Delta_z$. This is due to two combined effects. One is that satellites of galaxy pairs are not randomly distributed, but show a preference to inhabit the space between the galaxy pair, the so- called lopsided distribution found in \cite{2016ApJ...830..121L}. The second is that the primary galaxy pair is also separated along the line of sight. For a magnitude limit sample, more satellites would be observed around the primary closer to us. Now we find that there are 20.4\%  cent more satellites (with spectra redshift) around the close primary. In detail, the number fraction with $\Delta{z}$ > 0 is 0.64 for all sample (red solid line). The fraction with $\Delta{z}$ > 0 is 0.46 and 0.78 for the further primary (red dotted line) and close primary (red dashed line), respectively. So the total fraction for all sample is $(1.204*0.78+1*0.46)/(1.204+1) \sim 0.64$,  in agreement with what is shown in the red solid line. Therefore, the combined effect leads to a slightly skewed distribution towards positive $\Delta z$.

\begin{table}
\begin{center}
\begin{tabular}{c|c|c|c|c}
\hline
                     & All & $\rm < 100 \ kpc$ & $\rm < 200 \ kpc$ & $\rm < 300 \ kpc$  \\
\hline              
$\rm N_{total}$   & 1895075  &    5361 & 15700 & 30099  \\       
\hline
$\rm N_{spz}$   &  12665 (0.67\%) & 38 (0.71\%) & 120 (0.76\%) & 217 (0.72\%) \\
\hline
$\rm N_{phoz}$ & 1882410 & 5322 & 15580 & 29882 \\ 
\hline
\end{tabular}
\end{center}
\caption{The number of satellite with spectroscopic redshift and photometric redshift.}
\label{table:number}
\end{table}

\subsection{Alignment angles}
\label{sec:angles}

Before we investigate the shape alignment of satellite galaxy around the primary galaxy pair, it is useful to show the spatial schematic of the satellite-primary system in Fig.~\ref{fig:schematic}. The two black points represent the two primaries (`$\rm P_{1}$' and `$\rm P_{2}$') of a given galaxy pair. Two satellites (`$\rm S_{a}$' and `$\rm S_{b}$') found within the searching radius $\rm R_{search}$, belong to primary `$\rm P_{2}$' are also shown. The distribution of the angle $\theta~\angle~\rm P_{1}\rm P_{2}\rm S_{a}$ or $\theta^{\prime}~\angle~\rm P_{1}\rm P_{2}\rm S_{b}$ (subtended between the line connecting the primaries and the position of the satellite) was studied in detail by \cite{2016ApJ...830..121L}. We follow their  definition stating that a satellite is $\facing$ the other primary if $\theta\leq90^{\circ}$ ( as in `$\rm S_{a}$') or is {\it facing away} from the other primary if $\theta^{\prime}\leq90^{\circ}$ (as in  `$\rm S_{b}$'). Note that when saying $\facing$ or $\away$ satellite, we do not care about which primary this satellite belongs to.  If the galaxy pair is located in a filament we can also calculate the angle between the filament spine and the $\myvect{P_{1}P_{2}}$, which is denoted as $\gamma$. From the filament finder \citep{2014MNRAS.438.3465T} and galaxy pair selection process \citep{2012MNRAS.427..428G,2013MNRAS.434.1838G}, we could measure the position angle of filament spine and the $\myvect{P_{1}P_{2}}$ in the sky. Because our filament and galaxy pair are close to each other, thus the angle $\gamma$ can be simply calculated using the difference of those two position angle. Noting that we also restrict the $\gamma$ into $[0^{\circ}, 90^{\circ}]$.

The main goal of this study is to characterize, to what extent, the orientation of the major axis of the satellite galaxies is aligned with their primary galaxy pair system, namely to examine the distribution of the alignment angle, $\beta$ or $\beta^{\prime}$. Note that the alignment angle $\beta (\beta^{\prime})$ is the angle between the orientation of projected major axis of the satellite and its position vector relative to its primary.

The significance of any non-uniform distribution seen in a histogram of angle $X$ must be quantified so that the strength of the alignment can be measured. To do so, we compare with a control sample constructed in the following manner. For each satellite in our sample, we keep its position fixed but randomly draw 10 000 random position angles from the full position angle distribution (shown in the panel (c) of the Fig.~\ref{fig:method}), which means that the distribution of random satellite galaxies should follow the distribution of panel (c) in the Fig.~\ref{fig:method}. The purpose of this is to avoid the possible impact of uneven distribution of the position angle. Also, as shown in Fig.~\ref{fig:test}, the contamination tests indicate that the results of the analysis are likely not substantially affected by this flaw. The alignment signal can then be expressed as
\begin{equation}
  P(X) = \frac{N(X)}{\langle N_{R}(X)\rangle},
\end{equation}
where $X$ is the alignment angle denoted either $\beta$ or $\beta^{\prime}$, $N(X)$ is the number of satellites in each bin of alignment angle, and $\langle N_{R}(X)\rangle$ is the average number of satellites in the same bin obtained from the random control sample. The strength (or weakness) of the alignment can be characterized by $\sigma$, the deviation between the estimated probability distribution $P(X)$ and the set of null-hypothesis of random position angle distribution $P_R(X)$, which is calculated in units of the standard deviation of $P_R(X)$ of the aforementioned 10,000 random samples. In the absence of any alignment, $P(X)=1$. If $P(X)>1$ appears at a low values of $X$ it means that the major axis of satellites tends to point toward their primary: this is thus a positive measurement of the radial alignment. Whereas if  $P(X)>1$ appears at a larger value of $X$ this indicates that satellite major axes tend to be perpendicular to their position vector, a more tangential alignment.

%------------------------------------------------------------------------------------------
\begin{figure}
\centering
\centerline{\includegraphics[width=0.5\textwidth]{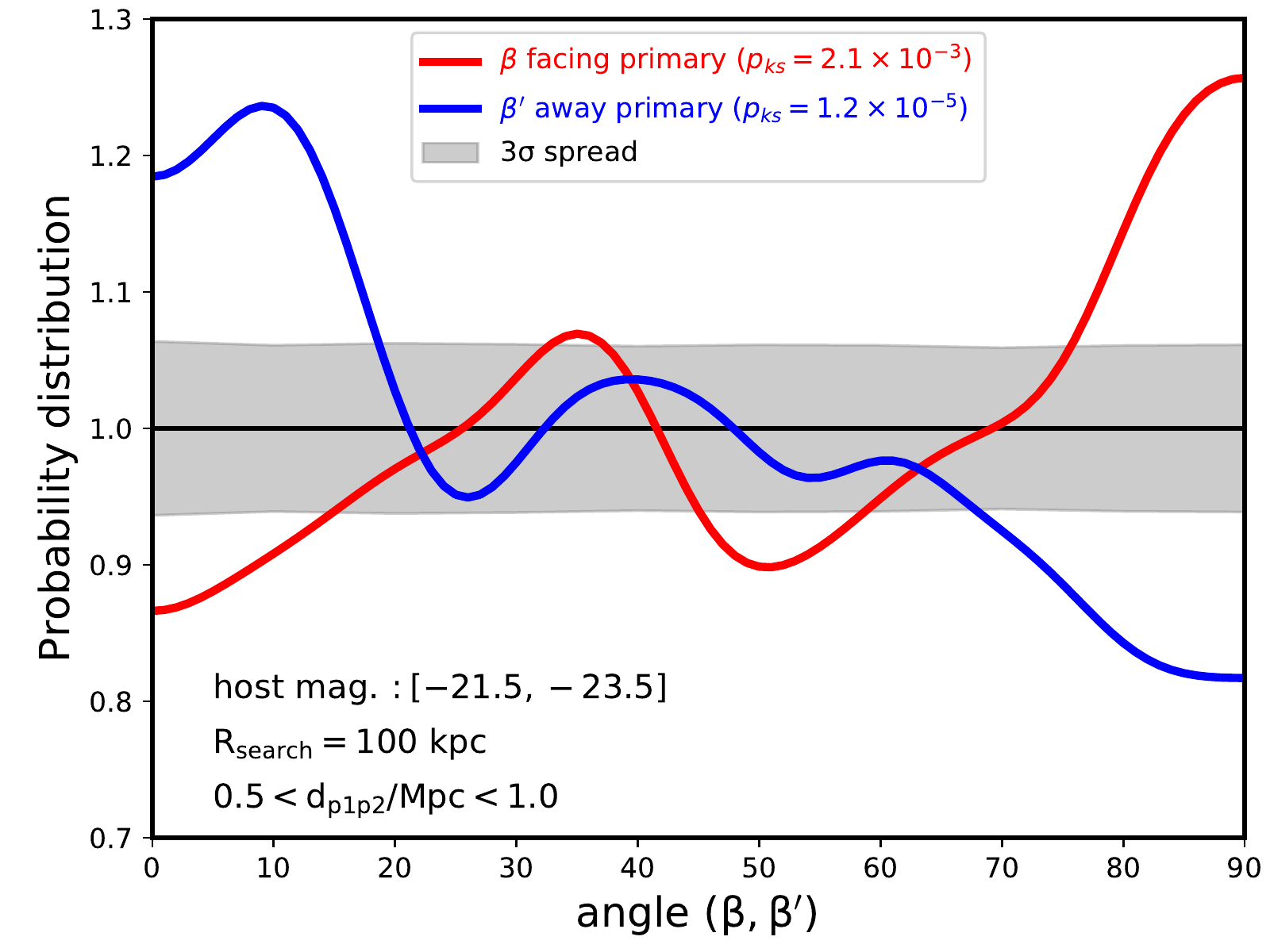}}
\caption{The normalized probability distribution of the angle, $\beta$ and $\beta^{\prime}$, between the orientation of the projected major axis of satellites and the position vector of the satellite-primary. Here we show the results from our fiducial sample. The host magnitude is within $[-21.5,-23.5]$ in $r$ band and satellites are selected with $\rm R_{search}<100 kpc$. The red curve represents $\beta$ (the angle for $\facing$ satellite facing the other primary) while the blue curve represents the distribution of $\beta^{\prime}$ (the angle for satellite facing $\away$ from the other primary). The grey band shows the spread of $3\sigma$ deviation from 10,000 random uniform distributions of the same size. The excess is seen at low value for $\beta^{\prime}$, and at high value for $\beta$.}
\label{fig:all}
\end{figure}
%------------------------------------------------------------------------------------------
\begin{figure}
\centering{\includegraphics[width=0.5\textwidth]{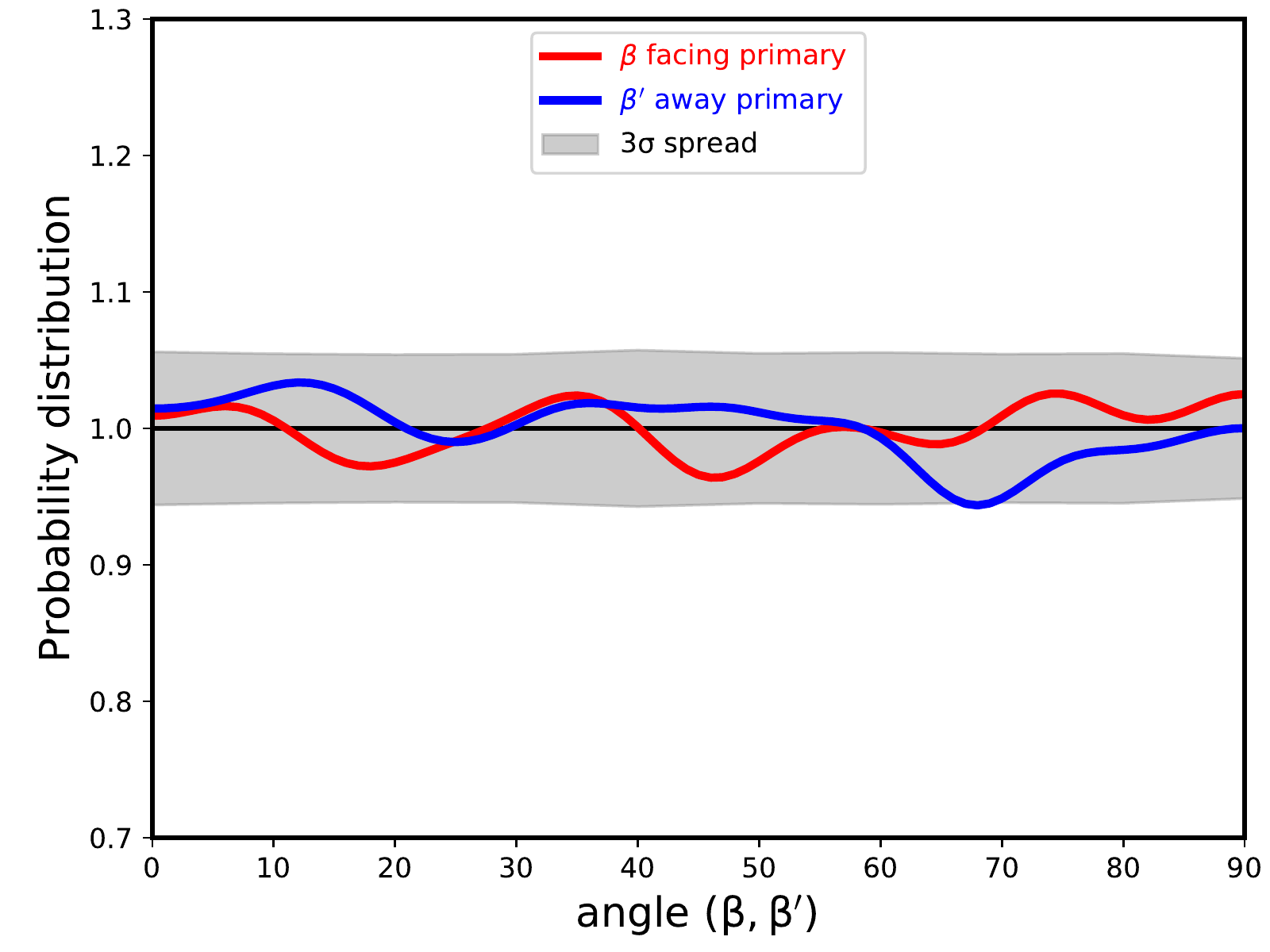}}
\centering{\includegraphics[width=0.5\textwidth]{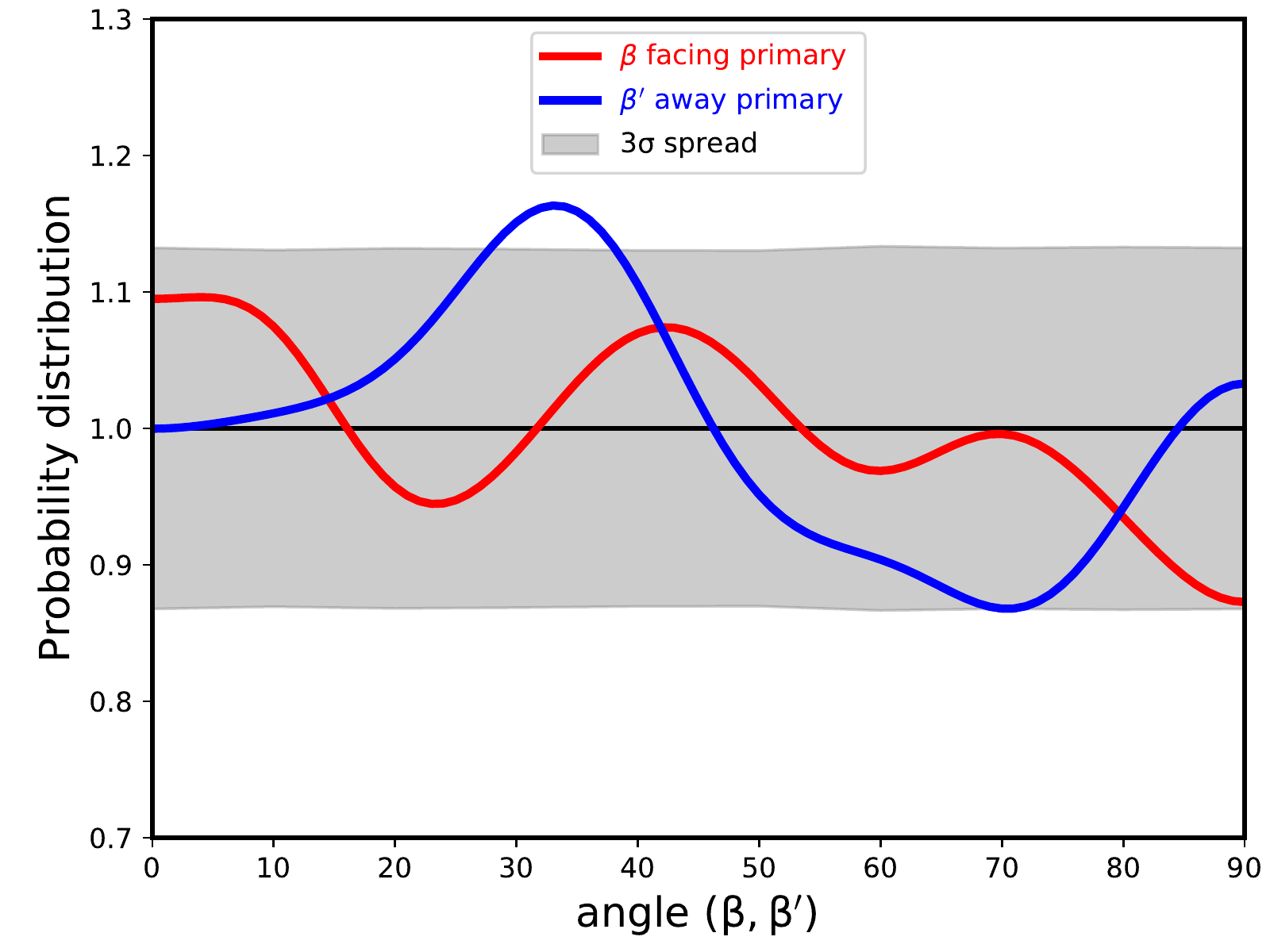}}
\caption{Similar as Fig.~\ref{fig:all} but for the alignment from background contamination. The top panel shows the shape alignment signals of all satellite galaxies within the searching radius but without photo-z cut. Lower panel is for signal from satellites around random primary galaxy pair (see text for more details).}
\label{fig:test}
\end{figure}
%------------------------------------------------------------------------------------------
\begin{figure*}
\centering
\centerline{\includegraphics[width=1.0\textwidth]{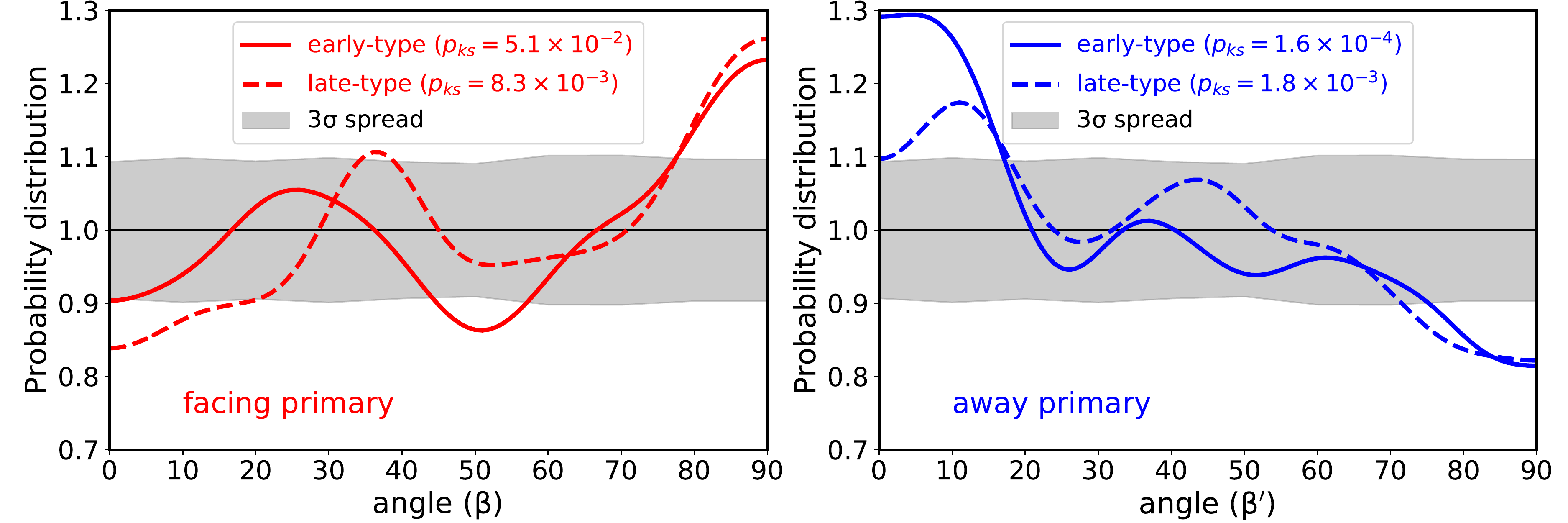}}
\caption{Same as Fig.~\ref{fig:all}, but for different morphology type of satellites. The results of `early-type' satellites and `late-type' satellites are shown as the solid and dashed lines, respectively.  The number of `early-type' satellites is roughly same with that of `late-type', which is reason for  relatively similar $3\sigma$ deviation and less significance.}
\label{fig:type}
\end{figure*}
%------------------------------------------------------------------------------------------

%------------------------------------------------------------------------------------------
\begin{figure*}
\centering
\centerline{\includegraphics[width=1.0\textwidth]{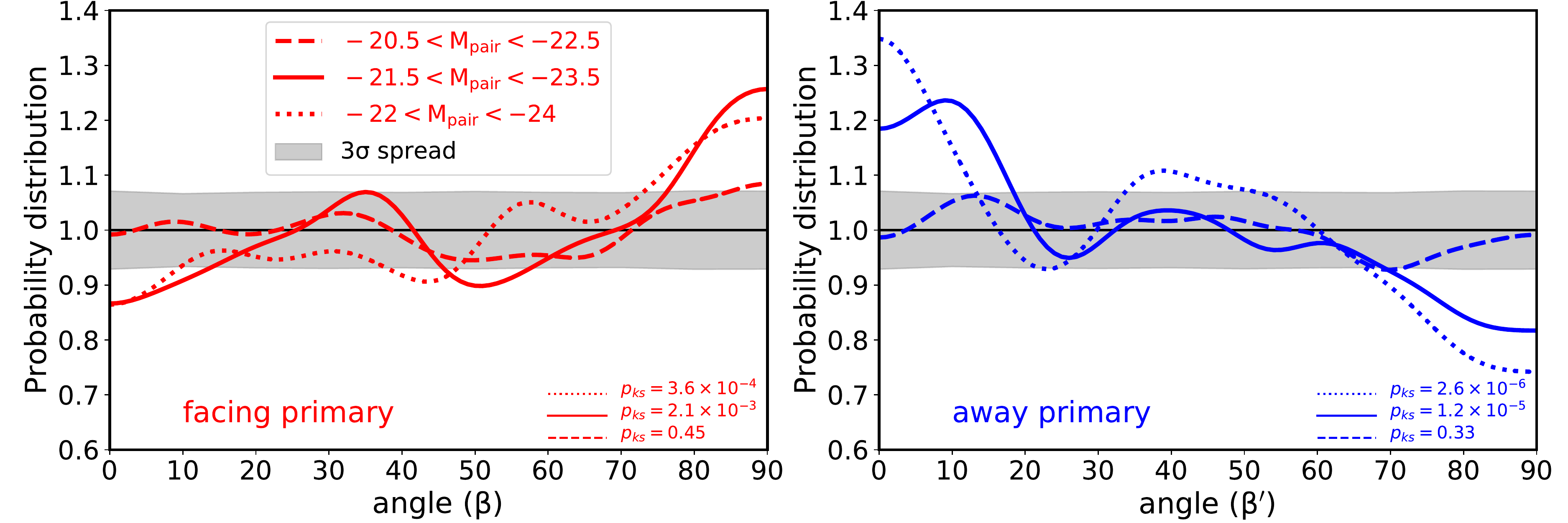}}
\caption{We examine the signal as a function of primary magnitude. The magnitudes of both two primary galaxies fall within the ranges $[-20.5, -22.5]$, $[-21.5, -23.5]$ and $[-22, -24]$, which are shown in different line styles. Note that the different magnitude between two primaries still falls within 1 mag and other parameters are kept fixed. Note that the grey bands show the spread of $3\sigma$ deviation of the sub-sample with the smallest sample size.}
\label{fig:hostmag}
\end{figure*}
%------------------------------------------------------------------------------------------

%%%%%%%%
%   Results
%%%%%%%%
\section{Results}
\label{sec:results}
In Fig.~\ref{fig:all}, we show the probability distribution of the alignment angle, $\beta$ or $\beta^{\prime}$, compared with a uniform distribution. The red line represents the result for satellite galaxies $\facing$ the other primary (namely $\beta$, see Fig.~\ref{fig:schematic}), and the blue line represents the result from satellite galaxies facing $\away$ from the other primary (namely $\beta^{\prime}$, see Fig.~\ref{fig:schematic}). The grey band shows the $3\sigma$ region expected from a uniform distribution of the same sample size. We can see that for the $\facing$  satellites (red line), the excess probability distribution above $3\sigma$ occurs at large value of $\beta \gtrsim76^{\circ}$. The excess reaches up to $\sim25\%$. A Kolmogorov-Smirnov (KS) test is performed to quantify the likelihood that these are consistent with being derived from an uniform distribution and it is indicated within Fig.~\ref{fig:all} as $p_{KS}=2.1\times10^{-3}$. So the uniform distribution hypothesis is rejected at the $\sim3\sigma$ level. This indicates that these $\facing$ satellites tend to be aligned tangentially to their primaries. For the $\away$ satellites, the result is the opposite. An excess above $3\sigma$ is seen at the low value of $\beta\ltrsim 18^{\circ}$, reaching an excess up to $\sim25\%$ with $p_{KS}=1.2\times10^{-5}$. This shows that, for these $\away$ satellite galaxies, their major axes are more inclined to point to the primary galaxies, i.e. radial alignment.

Now we check whether our results are significantly affected by background contamination. We quantify the background contami- nation in two ways. First, we fix the primary galaxy pair, but measure the alignment of all satellites within the projected searching radius (100 kpc), but without the photo-z cut. The results are shown in the top panel of Fig.~\ref{fig:test}, and demonstrate that there is a non-detectable alignment signal. Secondly, we construct a random sample for the primary galaxy pair by randomizing the orientation of the pair on the sky, keeping other parameters fixed, such as their redshifts, separation, and magnitude. We then use our previous criteria to select satellite galaxies (1200 satellites are now found). The results are shown in the lower panel of Fig.~\ref{fig:test}, which again shows that there is no detectable signal. Together, the two tests show that the background contamination will not affect our detected alignment signals in Fig.~\ref{fig:all}

In Fig.~\ref{fig:type}, we divide the satellite galaxies into two subsamples based on the value of their inverse concentration indices, $\mathcal{C}$. Early-type galaxies (with $\mathcal{C}<2.43$) are shown in solid lines while late-types (with $\mathcal{C}>2.43$) are shown in dashed lines. The number ratio of early-type satellite galaxies to late-type satellite galaxies is close to $1:1$, resulting in a similar spread of $3\sigma$ deviation for both samples.

In the $\facing$ region, shown in the left panel, it can be seen that the tangential distribution occur for both early- (red solid line) and late-type (red dashed line) satellite galaxies. Late-type satellite has a slightly stronger signal with an excess of $\sim27\%$ with $p_{KS}=8.3\times10^{-3}$ of the KS test, while for early-type satellites with relative smaller excess of $\sim~22\%$ and weaker confidence level with $p_{KS}=5.1\times10^{-2}$ of the KS test. However, in the $\away$ region (shown in the right panel), the results for early-type and late-type is opposite; early-type satellites show slightly stronger signal with an excess up to $\sim29\%$ with $p_{KS}=1.6\times10^{-4}$, while late-type satellites show relative weak signal of a $\sim18\%$ excess.

%------------------------------------------------------------------------------------------
\begin{figure*}
\centering
\centerline{\includegraphics[width=1.0\textwidth]{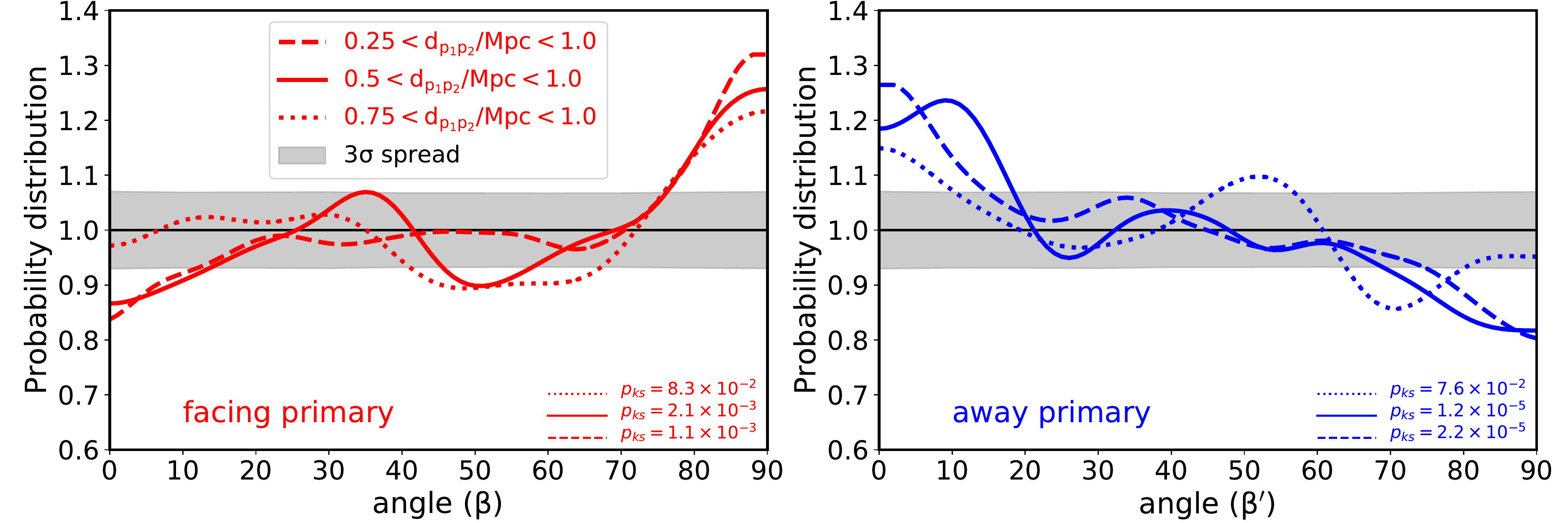}}
\caption{The signal is examined as a function of the projected separation between two member galaxies in galaxy pair. Galaxy pairs are divided into three bins, $\rm [0.25, 1.0] \ Mpc$ (dashed lines), $\rm [0.5,1.0] \ Mpc$ (fiducial sample, solid lines) and $\rm [0.75,1.0]\ Mpc$ (dotted lines), according to their projected separation. Note that the grey bands show the spread of $3\sigma$ deviation of the sub-sample with the smallest sample size.}
\label{fig:sep}
\end{figure*}
%------------------------------------------------------------------------------------------
%------------------------------------------------------------------------------------------
\begin{figure*}
\centerline{\includegraphics[width=1.0\textwidth]{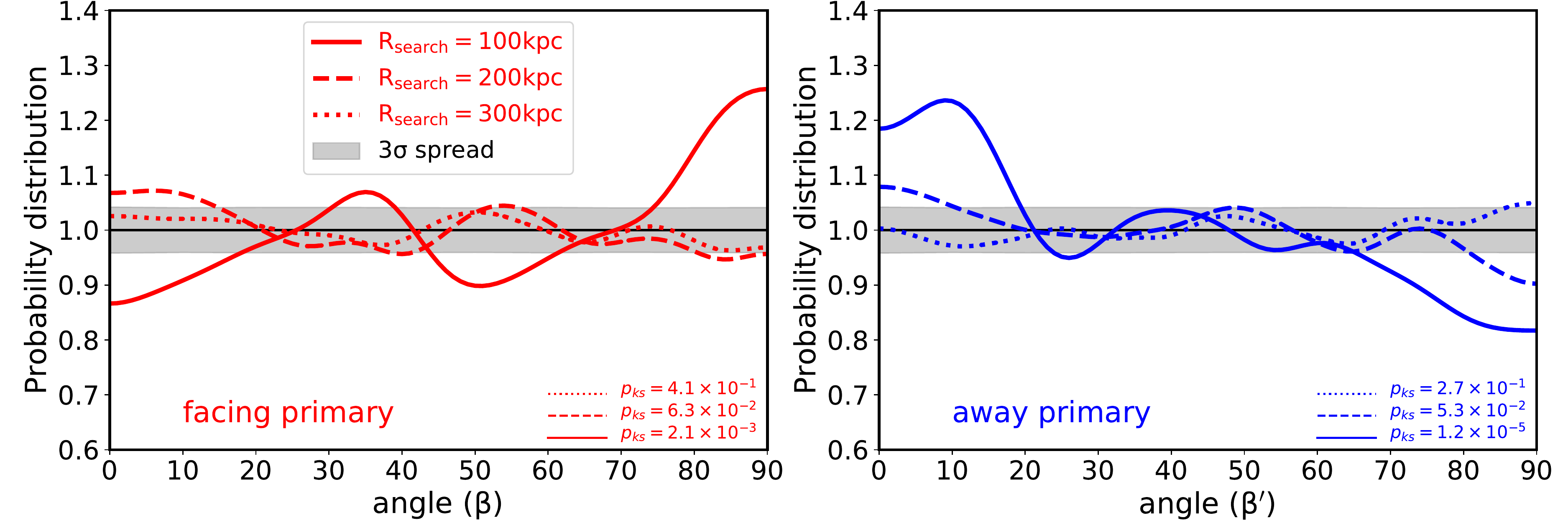}}
\caption{The signals for all satellites is examined as a function of search radii, $\rm R_{search}$. The results for the radius of $\rm100 \ kpc$, $\rm 200 \ kpc$ and $\rm 300 \ kpc$, are shown in different line styles respectively. Note that the grey bands show the spread of $3\sigma$ deviation of the sub-sample with the smallest sample size.}
\label{fig:searchR}
\end{figure*}
%------------------------------------------------------------------------------------------

Fig.~\ref{fig:all} and Fig.~\ref{fig:type} allow us to conclude that satellite galaxies {\it facing} their partner tend to have their major axes perpendicularly aligned to their position vectors. On the other hand,  satellite galaxies {\it facing away} from their partner tend to have their major axis pointing towards their primary host. A slightly stronger signal is seen in {\it facing away} region for early-type satellites, while similar signals are found both for early- and late-types {\it facing} their partners.

%------------------------------------------------------------------------------------------
\begin{figure*}
\centering
\centerline{\includegraphics[width=1.0\textwidth]{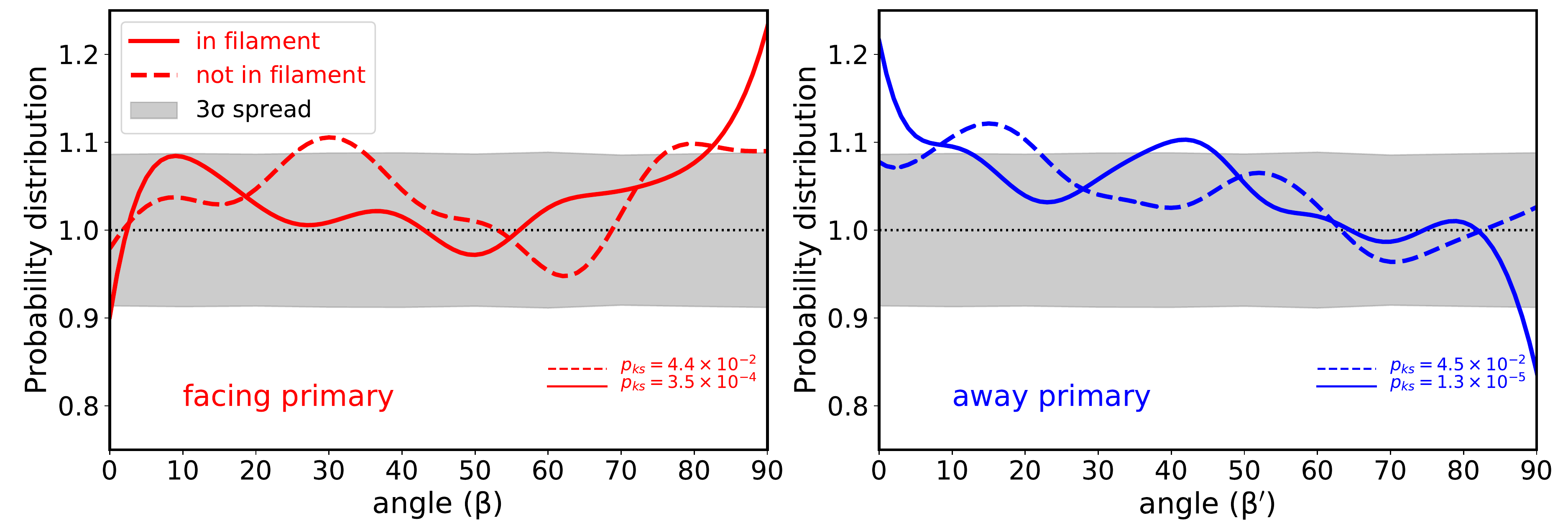}}
\caption{The impact of the large scale filament environments on the signal is examined. Our fiducial galaxy pairs are divided into two sub-samples: in the filament (solid lines) and not in the filament (dashed lines). The numbers of satellites in filament are slightly larger then not in filament same  resulting similar $3\sigma$ spread. Pairs in filament show more stronger and significance signal than that not in filament.}
\label{fig:filament}
\end{figure*}
%------------------------------------------------------------------------------------------

%------------------------------------------------------------------------------------------
\begin{figure*}
\centering
\centerline{\includegraphics[width=1.0\textwidth]{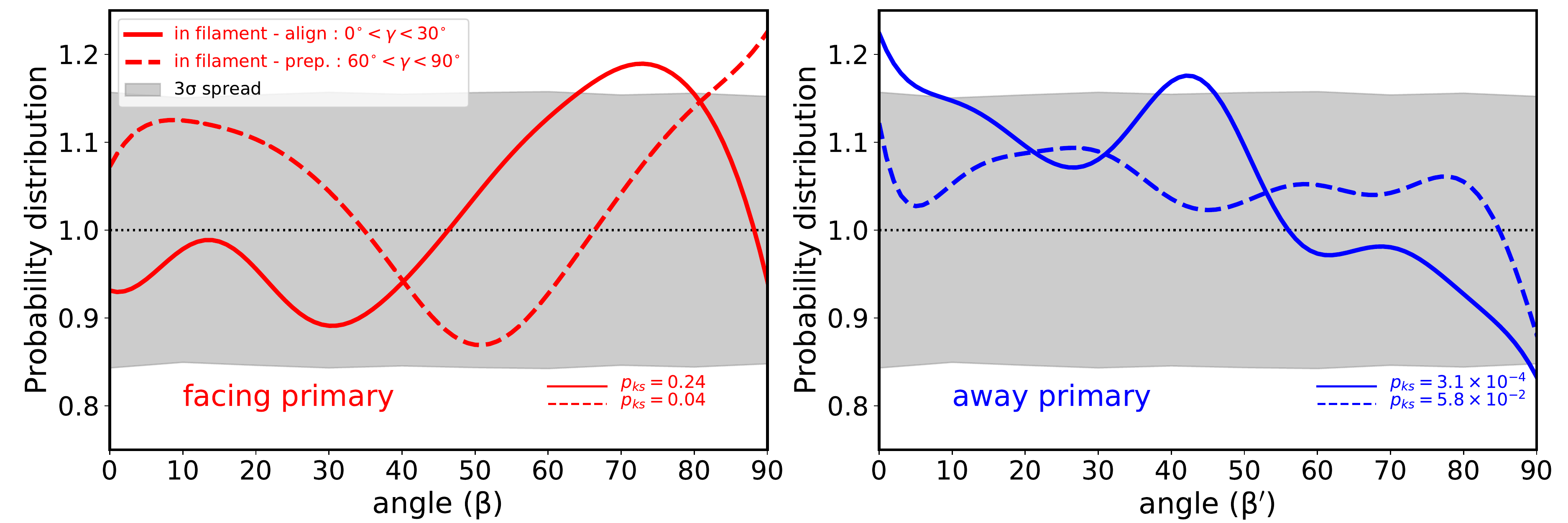}}
\caption{The signal is examined as a function of the angle between filament spine and the connecting line of two primaries, $\gamma$. In the left panel, we show pairs which tend to aligned ($0^{\circ}<\gamma<30^{\circ}$) with filament spine, and in the right panel, we show perpendicular case ($60^{\circ}<\gamma<60^{\circ}$). We note that for $30^{\circ}<\gamma<60^{\circ}$ is also checked, while all two lines always remain within $3\sigma$. Note that the grey bands show the spread of $3\sigma$ deviation of the sub-sample with the smallest sample size.}
\label{fig:filament_angle}
\end{figure*}
%------------------------------------------------------------------------------------------

\subsection{Dependence of the alignment}
In the following analysis, we systematically examine the influence of the host magnitude, projected host separation, and searching radius on the alignment signal. Note that when we study the effect of one parameter, the other parameters remain unchanged.

\subsubsection{Host magnitude}
\label{sec:mag}
In the fiducial sample, the differences of the magnitude between two primary galaxies are restricted to be within 1 mag, and magnitudes of both primaries are restricted in the range of $[-21.5, -23.5]$. In Fig.~\ref{fig:hostmag}, wwe examine how the primary galaxy brightness affects the radial and tangential signals seen above, when the other parameters are kept fixed. We divide the magnitudes of the primary galaxies into three bins, namely $[-20.5, -22.5]$, $[-21.5, -23.5]$ and $[-22, -24]$, corresponding to different line styles in Fig.~\ref{fig:hostmag}. 

It can be seen that the dependence on the host magnitude for the tangential alignment in the $\facing$ region (left-hand panel) is not particularly obvious. Satellite galaxies of the primaries with the lowest magnitude bin (red dashed line) show a non-detectable signal and the probability distributions always lie well within the $3\sigma$ interval expected from a uniform distribution. Signals of tangential alignment are marginally seen for the intermediate and highest magnitudes (solid and dotted lines). In the $\away$ region (right- hand panel), the signal shows a slightly stronger dependence on magnitude. In the lowest magnitude bin, the probability distributions are also within $3\sigma$ interval. {\it The KS test indicates that the strength of radial alignment gradually increases with the brightness of primary galaxies.}.

\subsubsection{Host separation}
\label{sec:dis}
Another parameter that can potentially affect the radial and tangential alignment of satellite galaxies is the separation between two member galaxies in galaxy pairs. The smaller the separation between the primaries, the stronger is the gravitational and tidal force experienced by their satellite galaxies. Hence, in this section, we investigate the effect of the galaxy pair separation on the signal strength. The distribution of pair separation is shown as a red solid line in Fig.~\ref{fig:method}d. We divide the galaxy pairs into three bins with projected separation within $[0.25, 1.0]$, $[0.5,1.0]$ and $[0.75, 1.0]$, in units of $\rm Mpc$. Note that the bin of $[0.5,1.0]$ is our fiducial sample.

In Fig.~\ref{fig:sep}, we show the effect of varying the (projected) separation of the primary galaxy pair. Overall, we do see the signal for all primary galaxy pairs with different distances, but the dependence on pair separation is not very strong. The KS test shows a slight dependence that the signal increases with decreasing pair separation. As in this work we only focus on primary galaxy pairs with a luminosity range similar to Local Group, the sample size is not large enough to quantify the signal more accurately. It would be interesting to investigate the dependence on pair separation from larger sample size by imposing a wide luminosity range.

\subsubsection{Radius to search satellites}
\label{sec:searchR}

In Fig.~\ref{fig:searchR}, we show the radial and tangential alignment of the satellite galaxies as a function of the searching radius. We examined three searching radii of $100 \kpc$, $200 \kpc$, and $300 \kpc$. The solid lines are the results for our fiducial searching radius, the dashed lines are for the radius of $200 \kpc$, and dotted lines are for the radius of $300 \kpc$. Overall, the strength of the signal increases with the decreasing of radius. What is more interesting is that for $\facing$ satellite galaxies (left-hand panel), its signal gradually changes into a tangential alignment from a radial alignment as the searching radius decreases. For satellite galaxies in the $\away$ area, their radial alignment signals gradually become random as the searching radius increases.

\subsection{Filament environment}
We further examine if the radial and tangential signals found above are affected by the presence of a cosmic filament. We divide our fiducial sample into two subsamples according to the filament classification of the pair, where the details of the filament catalogue can be found inSection.~\ref{sec:cata} and \cite{2014MNRAS.438.3465T}.

In Fig.~\ref{fig:filament}, we show the alignment angle of satellites whose hosts in a filament (Fig.~\ref{fig:filament}, solid line) or not in a filament (Fig.~\ref{fig:filament}, dashed lines). Note that the number of satellites in the filament is slightly larger than that not in the filament \citep{2015ApJ...800..112G}.  In the left-hand (right-hand) panel, we show the results for the cases of $\facing$ ($\away$) satellites when the pairs are either in or not in filaments. Galaxy pairs found in filaments show a significantly stronger signal than those pairs not in filamentary environments. Up to  $\sim22\%$ more satellites than expected from a random distribution are found radially aligned (for satellites facing away, left-hand panel) and tangentially aligned (for satellites facing their partner, right-hand panel).

\cite{2015A&A...576L...5T} howed that galaxy pairs are aligned with galactic filaments. To analyse whether satellite alignment depends on the pair alignment with filament, we divide satellites into two subsamples with the connecting vector either aligned with or perpendicular to their host filament. This is done according to the angle $\gamma$ (see Section.~\ref{sec:angles} for the definition) between the direction of the filament and the line connecting the two primaries. In Fig.~\ref{fig:filament_angle}, we show the distribution of the alignment angle $\beta$ for satellites whose host is either aligned with ( $\gamma < 30^{\circ}$) or perpendicular to ($\gamma > 60^{\circ}$) the host filament (the definition of angle $\gamma$ can be found in Section. ~\ref{sec:angles}). Note that due to the relative smaller number of pairs in this sub-sub division, the $3\sigma$  standard deviation of a random sample is relatively wide. 

Interestingly, the signal seen in the in-filament sample where the pair is aligned with the filament (namely satellites facing the other primary tend to be tangentially aligned while satellites facing away from their partner tend to be radially aligned) echoes the full fiducial sample, albeit at low statistical significance. Pairs of galaxies that are perpendicularly aligned to the host filament only show the tangentially aligned satellites for those facing their partner. That said, mostly due to the decreased sample size the further sub- sub divisions shown in Fig.~\ref{fig:filament_angle} the $3\sigma$ brand is wide, barring us from making any strong claims on the results. It should be noted that we also tested samples where the hosts were neither aligned parallel or perpendicular to the filament (namely $30^{\circ}<\gamma <60^{\circ}$) and found that both facing and facing away lines distributions always remain within $3\sigma$.

%%=====================================================================

%%%%%%%%%%%%%%%%%
%   Summary and Discussion
%%%%%%%%%%%%%%%%%
%------------------------------------------------------------------------------------------
\begin{figure}
\centerline{\includegraphics[width=0.5\textwidth]{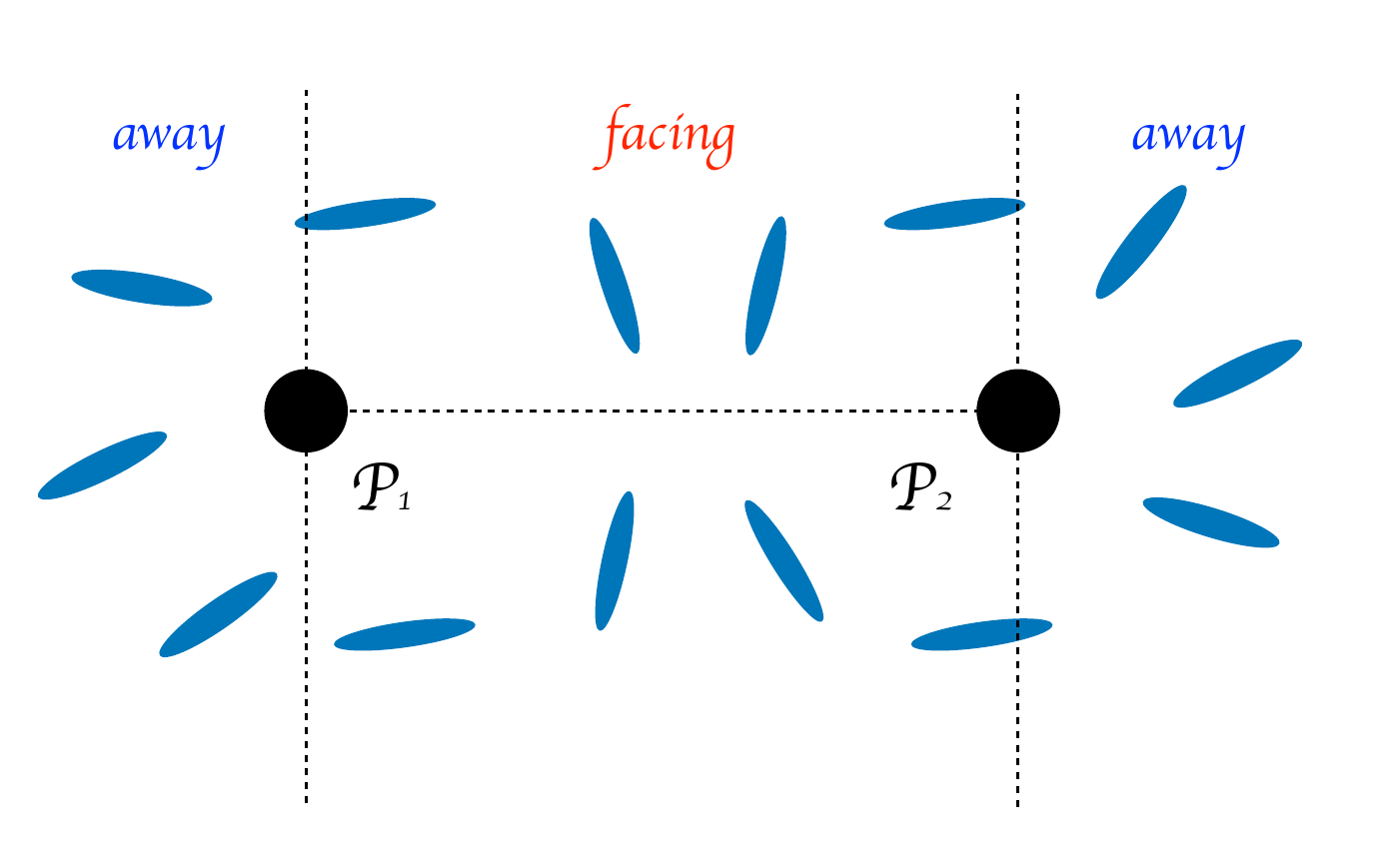}}
\caption{A cartoon figure to depict the orientation of satellite galaxies in galaxy pair system based on our results. The little blue ellipses represent satellites, and direction of the long axis of each ellipse represents the orientation of the projected major axis of the satellites. In regions where the satellite galaxies are facing $\away$ from the other primary, the major axis of the satellites tends to preferentially point towards their primary. While in region $\facing$ the other primary, the region between two primaries, satellite major axes tend to be tangentially distributed.}
\label{fig:cartoon}
\end{figure}
%------------------------------------------------------------------------------------------

\section{Summary and Discussion}
Using catalogue constructed from SDSS DR13, we investigate the shape alignment of satellites in galaxy pair systems. We focus on the alignment between the direction of the projected major axis of satellites and the position vector of satellite with respect to its host. The selection criterion of the galaxy pairs is motivated by the Local Group: we restricted the magnitude of our selected galaxy pairs to be within $[-21.5, -23.5]$, their projected separation to be between $[0.5, 1.0]$ $\mpc$, and satellites are selected to be within $100\kpc$ of their host (see our fiducial sample selection in Section~\ref{sec:method}).

Our main results of the fiducial sample can be summed up as follows (also see Fig.~\ref{fig:all}): {\it the orientation of the satellite galaxy is correlated with its location with respect to the primary galaxy pair}. Specifically, satellites that are found in between the pair tend to align their major axes perpendicular to their position vector, while those satellites that are further away from the other pair member tend to have their major axes pointing towards their host. For an intuitive understanding, we show the spatial depiction in Fig.~\ref{fig:cartoon}.  This figure explains the results we obtained that satellites located in the `facing' region show projected major axis preferentially perpendicular to the connecting line between primaries, while satellites that are located in the `away' region show a radial alignment trend.

We further examined the influence of satellite galaxy morphology on the signal. There is no strong dependence while it found that the signals of tangential distribution are slightly stronger for late-type satellites (solid lines in Fig.~\ref{fig:type}), whereas for early-type satellites, the signals of radial alignment are stronger (dashed line in Fig.~\ref{fig:type}). In order to determine whether the alignment signal depends on the specific choice of parameters used to identify our fiducial sample, a systematic study was conducted in which the results of varying host magnitude, pair separation, and searching radius are examined. We found that, in general, the strength of the alignment signal increases with the brightness of the hosts (Fig.~\ref{fig:hostmag}), while it decreases with increasing of the pair separation (Fig.~\ref{fig:sep}) and searching radius (Fig.~\ref{fig:searchR}).

In addition, we also examined the impact of a  large scale filament on the alignment signal. Using the filament catalog from \cite{2014A&A...566A...1T}, it is found that for satellites in filament environments facing $\away$ from the primary galaxy, the alignment signal is found to be stronger when pairs are aligned with the filament direction, while being weaker when pairs are perpendicular to the filament direction (right-hand panel in Fig.~\ref{fig:filament_angle}). For satellites in filament environments $\facing$ their host galaxy, the result is opposite (left-hand panel in Fig.~\ref{fig:filament_angle}).

Most existing studies have focused on the shape alignment of satellites in isolated systems, especially in galaxy groups or clusters. Though current studies have not reached a consensus whether the orientation of satellite galaxies is random or radially aligned with the central galaxy \citep[e.g.,][and references therein]{2018arXiv180802573P, 2018MNRAS.474.4772H}, a radial alignment is more consistent with the early theoretical predictions \citep{1994MNRAS.270..390C, 2007ApJ...671.1135K, 2008ApJ...672..825P}, in which the direction of the major axis of satellite galaxies is correlated with the tidal force from the host galaxy \citep{2008ApJ...672..825P}. In this work we only focus on satellites in Local Group-like pairs, and split the satellites into classes depending on where they are located with respect to the geometry of the pair (see Fig.~\ref{fig:cartoon}). Our results are qualitatively consistent with the theoretical prediction. For satellites in the `away' region, the satellites are often more close to one of the primary galaxies, thus the tidal force of the nearby primary galaxy dominates their orientation and we observe a radial alignment signal, similar to that seen in an isolated galaxy-scale system \citep[e.g.,][]{2006ApJ...644L..25A}. For satellites in the $\facing$ area (between the pair), the satellites will be influenced by the tidal force from both primary galaxies in the pair, so the radial alignment is changed to tangential alignment.

\cite{2017MNRAS.472.2670S} provide evidence to show that the major axes of the Milky Way satellites lie within a preferential plane, while no signal was found in M31. Our results may provide insight into understanding the shape alignment of satellites in Local Group. We provide a possible idea for studying the LG satellite galaxy shape alignment, and look for the tangential alignment in the `facing' region and the radial alignment in the 'away' region.

Our results may have an implication on how satellite galaxies are accreted in galaxy pair system. Many related studies \citep[e.g.,][]{2011MNRAS.411.1525L, 2014MNRAS.443.1274L, 2014ApJ...786....8W, 2015ApJ...807...37S,2015ApJ...813....6K, 2017MNRAS.468L.123W, 2018MNRAS.473.1562W} pointed out that satellite galaxies tend to be accreted along or perpendicular to the filamentary direction depending on the mass of the host galaxy. Most of these studies do not distinguish whether the host galaxies are isolated or in pairs, which undoubtedly has  impacts the satellite galaxy's accretion history and dynamics. Furthermore, satellite galaxies are mainly accreted along the filament direction \citep{2014MNRAS.443.1274L, 2014ApJ...786....8W, 2015ApJ...807...37S, 2015ApJ...813....6K, 2017MNRAS.468L.123W, 2018MNRAS.473.1562W}. For pairs that tend to align with the filament direction, its `away' area happens to be at the front end of the mass flow within a filament, while for pairs that tend to be perpendicular to the filament direction, the `facing' area is right at the front of the mass flow. The dynamically active of the primary pairs may contribute to the difference in their signals, which requires further investigation.

Finally, the shape alignment in galaxy pairs is also relevant to the issue of intrinsic alignment in weak gravitational lensing. In the context of weak lensing, a coherent distortion of background galaxy images can be used to constrain cosmological parameters \citep[e.g.,][]{2013MNRAS.432.2433H, 2017MNRAS.465.1454H}. However, one of the significant systematic contributions comes from the intrinsic galaxy shape alignment. As discussed in \cite{2018ApJ...853...25W}, if satellite galaxies are perfectly radially aligned with their host galaxies, the shear correlation on cluster scales will be strongly enhanced and can be ruled out by the data. If we consider satellite galaxies of pair system, the satellites in the ‘facing’ region, which contribute a tangential alignment, will dilute the radial alignment signal expected in an isolated system. Thus the shape alignment of satellites is more complicated than the simple models that only consider the galaxy shape alignment in isolated haloes \citep[e.g.,][]{2010MNRAS.402.2127S} and deserve more investigation in detail.

\section{Acknowledgments}
We thank the referee for careful reading and constructive sug- gestions that improved the presentation of our paper. The work is supported by the NSFC (Nos. 11333008, 11825303, 11861131006, 11703091), the 973 program (No. 2015CB857003, No. 2013CB834900). QG acknowledges support by the NSFC (No. 11743003). NIL acknowledges financial support of the Project IDEXLYON at the University of Lyon under the Investments for the Future Program (ANR-16-IDEX-0005). ET was supported by ETAg grants (IUT40-2, IUT26-2) and by EU through the ERDF CoE grants TK133 and MOBTP86.

%%%%%%%%%%%%%%%
%.         Bibliography
%%%%%%%%%%%%%%%

{}

%%%%%%%%%%%%%
%       Appendices
%%%%%%%%%%%%%

\label{lastpage}

\begin{thebibliography}{}

%Retrieved 125 abstracts, starting with number 1.  Total number selected: 126.

\bibitem[\protect\citeauthoryear{Agustsson \& Brainerd}{2010}]{2010ApJ...709.1321A} Agustsson I., Brainerd T.~G., 2010, ApJ, 709, 1321 
\bibitem[\protect\citeauthoryear{Agustsson \& Brainerd}{2006}]{2006ApJ...650..550A} Agustsson I., Brainerd T.~G., 2006, ApJ, 650, 550 
\bibitem[\protect\citeauthoryear{Agustsson \& Brainerd}{2006}]{2006ApJ...644L..25A} Agustsson I., Brainerd T.~G., 2006, ApJ, 644, L25 
\bibitem[\protect\citeauthoryear{Albareti et al.}{2017}]{2017ApJS..233...25A} Albareti F.~D., et al., 2017, ApJS, 233, 25 
\bibitem[\protect\citeauthoryear{Alonso et al.}{2006}]{2006RMxAC..26..187A} Alonso M.~S., Tissera P.~B., Lambas D.~G., Coldwell G., 2006, RMxAC, 26, 187 
\bibitem[\protect\citeauthoryear{Alonso et al.}{2007}]{2007MNRAS.375.1017A} Alonso M.~S., Lambas D.~G., Tissera P., Coldwell G., 2007, MNRAS, 375, 1017 
\bibitem[\protect\citeauthoryear{Alonso et al.}{2004}]{2004MNRAS.352.1081A} Alonso M.~S., Tissera P.~B., Coldwell G., Lambas D.~G., 2004, MNRAS, 352, 1081 
\bibitem[\protect\citeauthoryear{Alonso et al.}{2012}]{2012A&A...539A..46A} Alonso S., Mesa V., Padilla N., Lambas D.~G., 2012, A\&A, 539, A46 
\bibitem[\protect\citeauthoryear{Angulo et al.}{2009}]{2009MNRAS.399..983A} Angulo R.~E., Lacey C.~G., Baugh C.~M., Frenk C.~S., 2009, MNRAS, 399, 983 
\bibitem[\protect\citeauthoryear{Angus et al.}{2016}]{2016MNRAS.462.3221A} Angus G.~W., Coppin P., Gentile G., Diaferio A., 2016, MNRAS, 462, 3221 
\bibitem[\protect\citeauthoryear{Arag{\'o}n-Calvo et al.}{2007}]{2007ApJ...655L...5A} Arag{\'o}n-Calvo M.~A., van de Weygaert R., Jones B.~J.~T., van der Hulst J.~M., 2007, ApJ, 655, L5 
\bibitem[\protect\citeauthoryear{Aubert, Pichon, \& Colombi}{2004}]{2004MNRAS.352..376A} Aubert D., Pichon C., Colombi S., 2004, MNRAS, 352, 376 
\bibitem[\protect\citeauthoryear{Azzaro et al.}{2007}]{2007MNRAS.376L..43A} Azzaro M., Patiri S.~G., Prada F., Zentner A.~R., 2007, MNRAS, 376, L43 
\bibitem[\protect\citeauthoryear{Bailin et al.}{2008}]{2008MNRAS.390.1133B} Bailin J., Power C., Norberg P., Zaritsky D., Gibson B.~K., 2008, MNRAS, 390, 1133 
\bibitem[\protect\citeauthoryear{Barber et al.}{2015}]{2015MNRAS.447.1112B} Barber C., Starkenburg E., Navarro J.~F., McConnachie A.~W., 2015, MNRAS, 447, 1112 
\bibitem[\protect\citeauthoryear{Bardeen et al.}{1986}]{1986ApJ...304...15B} Bardeen J.~M., Bond J.~R., Kaiser N., Szalay A.~S., 1986, ApJ, 304, 15 
\bibitem[\protect\citeauthoryear{Barton, Geller, \& Kenyon}{2000}]{2000ApJ...530..660B} Barton E.~J., Geller M.~J., Kenyon S.~J., 2000, ApJ, 530, 660 
\bibitem[\protect\citeauthoryear{Bowden, Evans, \& Belokurov}{2014}]{2014ApJ...793L..42B} Bowden A., Evans N.~W., Belokurov V., 2014, ApJ, 793, L42 
\bibitem[\protect\citeauthoryear{Brainerd \& Specian}{2003}]{2003ApJ...593L...7B} Brainerd T.~G., Specian M.~A., 2003, ApJ, 593, L7 
\bibitem[\protect\citeauthoryear{Cautun et al.}{2015}]{2015MNRAS.452.3838C} Cautun M., Bose S., Frenk C.~S., Guo Q., Han J., Hellwing W.~A., Sawala T., Wang W., 2015, MNRAS, 452, 3838 
\bibitem[\protect\citeauthoryear{Chisari et al.}{2014}]{2014MNRAS.445..726C} Chisari N.~E., Mandelbaum R., Strauss M.~A., Huff E.~M., Bahcall N.~A., 2014, MNRAS, 445, 726 
\bibitem[\protect\citeauthoryear{Ciotti \& Dutta}{1994}]{1994MNRAS.270..390C} Ciotti L., Dutta S.~N., 1994, MNRAS, 270, 390 
\bibitem[\protect\citeauthoryear{Conn et al.}{2013}]{2013ApJ...766..120C} Conn A.~R., et al., 2013, ApJ, 766, 120 
\bibitem[\protect\citeauthoryear{Conroy et al.}{2007}]{2007ApJ...654..153C} Conroy C., et al., 2007, ApJ, 654, 153 
\bibitem[\protect\citeauthoryear{Deason et al.}{2011}]{2011MNRAS.415.2607D} Deason A.~J., et al., 2011, MNRAS, 415, 2607 
\bibitem[\protect\citeauthoryear{Drlica-Wagner et al.}{2015}]{2015ApJ...813..109D} Drlica-Wagner A., et al., 2015, ApJ, 813, 109 
\bibitem[\protect\citeauthoryear{Ellison et al.}{2013a}]{2013MNRAS.435.3627E} Ellison S.~L., Mendel J.~T., Patton D.~R., Scudder J.~M., 2013a, MNRAS, 435, 3627 
\bibitem[\protect\citeauthoryear{Ellison et al.}{2013b}]{2013MNRAS.430.3128E} Ellison S.~L., Mendel J.~T., Scudder J.~M., Patton D.~R., Palmer M.~J.~D., 2013b, MNRAS, 430, 3128 
\bibitem[\protect\citeauthoryear{Ellison, Patton, \& Hickox}{2015}]{2015MNRAS.451L..35E} Ellison S.~L., Patton D.~R., Hickox R.~C., 2015, MNRAS, 451, L35 
\bibitem[\protect\citeauthoryear{Ellison et al.}{2011}]{2011MNRAS.418.2043E} Ellison S.~L., Patton D.~R., Mendel J.~T., Scudder J.~M., 2011, MNRAS, 418, 2043 
\bibitem[\protect\citeauthoryear{Ellison et al.}{2008}]{2008AJ....135.1877E} Ellison S.~L., Patton D.~R., Simard L., McConnachie A.~W., 2008, AJ, 135, 1877 
\bibitem[\protect\citeauthoryear{Ellison et al.}{2010}]{2010MNRAS.407.1514E} Ellison S.~L., Patton D.~R., Simard L., McConnachie A.~W., Baldry I.~K., Mendel J.~T., 2010, MNRAS, 407, 1514 
\bibitem[\protect\citeauthoryear{Faltenbacher et al.}{2008}]{2008ApJ...675..146F} Faltenbacher A., Jing Y.~P., Li C., Mao S., Mo H.~J., Pasquali A., van den Bosch F.~C., 2008, ApJ, 675, 146 
\bibitem[\protect\citeauthoryear{Faltenbacher et al.}{2007}]{2007ApJ...662L..71F} Faltenbacher A., Li C., Mao S., van den Bosch F.~C., Yang X., Jing Y.~P., Pasquali A., Mo H.~J., 2007, ApJ, 662, L71 
\bibitem[\protect\citeauthoryear{Faltenbacher et al.}{2009}]{2009RAA.....9...41F} Faltenbacher A., Li C., White S.~D.~M., Jing Y.-P., Shu-DeMao, Wang J., 2009, RAA, 9, 41 
\bibitem[\protect\citeauthoryear{Ganeshaiah Veena et al.}{2018}]{2018MNRAS.481..414G} Ganeshaiah Veena P., Cautun M., van de Weygaert R., Tempel E., Jones B.~J.~T., Rieder S., Frenk C.~S., 2018, MNRAS, 481, 414
\bibitem[\protect\citeauthoryear{Guo et al.}{2012}]{2012MNRAS.427..428G} Guo Q., Cole S., Eke V., Frenk C., 2012, MNRAS, 427, 428 
\bibitem[\protect\citeauthoryear{Guo et al.}{2011}]{2011MNRAS.417..370G} Guo Q., Cole S., Eke V., Frenk C., 2011, MNRAS, 417, 370 
\bibitem[\protect\citeauthoryear{Guo et al.}{2013}]{2013MNRAS.434.1838G} Guo Q., Cole S., Eke V., Frenk C., Helly J., 2013, MNRAS, 434, 1838 
\bibitem[\protect\citeauthoryear{Guo, Tempel, \& Libeskind}{2015}]{2015ApJ...800..112G} Guo Q., Tempel E., Libeskind N.~I., 2015, ApJ, 800, 112 
\bibitem[\protect\citeauthoryear{Hammer et al.}{2013}]{2013MNRAS.431.3543H} Hammer F., Yang Y., Fouquet S., Pawlowski M.~S., Kroupa P., Puech M., Flores H., Wang J., 2013, MNRAS, 431, 3543 
\bibitem[\protect\citeauthoryear{Hawley \& Peebles}{1975}]{1975AJ.....80..477H} Hawley D.~L., Peebles P.~J.~E., 1975, AJ, 80, 477 
\bibitem[\protect\citeauthoryear{Heymans et al.}{2013}]{2013MNRAS.432.2433H} Heymans C., et al., 2013, MNRAS, 432, 2433 
\bibitem[\protect\citeauthoryear{Hildebrandt et al.}{2017}]{2017MNRAS.465.1454H} Hildebrandt H., et al., 2017, MNRAS, 465, 1454 
\bibitem[\protect\citeauthoryear{Holmberg}{1969}]{1969ArA.....5..305H} Holmberg E., 1969, ArA, 5, 305 
\bibitem[\protect\citeauthoryear{Huang et al.}{2018}]{2018MNRAS.474.4772H} Huang H.-J., Mandelbaum R., Freeman P.~E., Chen Y.-C., Rozo E., Rykoff E., 2018, MNRAS, 474, 4772 
\bibitem[\protect\citeauthoryear{Hung \& Ebeling}{2012}]{2012MNRAS.421.3229H} Hung C.-L., Ebeling H., 2012, MNRAS, 421, 3229 
\bibitem[\protect\citeauthoryear{Ibata et al.}{2013}]{2013Natur.493...62I} Ibata R.~A., et al., 2013, Natur, 493, 62 
\bibitem[\protect\citeauthoryear{Jing \& Suto}{2002}]{2002ApJ...574..538J} Jing Y.~P., Suto Y., 2002, ApJ, 574, 538 
\bibitem[\protect\citeauthoryear{Joachimi et al.}{2015}]{2015SSRv..193....1J} Joachimi B., et al., 2015, SSRv, 193, 1 
\bibitem[\protect\citeauthoryear{Kang et al.}{2005}]{2005A&A...437..383K} Kang X., Mao S., Gao L., Jing Y.~P., 2005, A\&A, 437, 383 
\bibitem[\protect\citeauthoryear{Kang et al.}{2007}]{2007MNRAS.378.1531K} Kang X., van den Bosch F.~C., Yang X., Mao S., Mo H.~J., Li C., Jing Y.~P., 2007, MNRAS, 378, 1531 
\bibitem[\protect\citeauthoryear{Kang \& Wang}{2015}]{2015ApJ...813....6K} Kang X., Wang P., 2015, ApJ, 813, 6 
\bibitem[\protect\citeauthoryear{Karachentsev \& Kashibadze}{2006}]{2006Ap.....49....3K} Karachentsev I.~D., Kashibadze O.~G., 2006, Ap, 49, 3 
\bibitem[\protect\citeauthoryear{Kennicutt et al.}{1987}]{1987AJ.....93.1011K} Kennicutt R.~C., Jr., Keel W.~C., van der Hulst J.~M., Hummel E., Roettiger K.~A., 1987, AJ, 93, 1011
\bibitem[\protect\citeauthoryear{Knebe et al.}{2008a}]{2008MNRAS.386L..52K} Knebe A., Draganova N., Power C., Yepes G., Hoffman Y., Gottl{\"o}ber S., Gibson B.~K., 2008a, MNRAS, 386, L52 
\bibitem[\protect\citeauthoryear{Knebe et al.}{2004}]{2004ApJ...603....7K} Knebe A., Gill S.~P.~D., Gibson B.~K., Lewis G.~F., Ibata R.~A., Dopita M.~A., 2004, ApJ, 603, 7 
\bibitem[\protect\citeauthoryear{Knebe et al.}{2010}]{2010MNRAS.405.1119K} Knebe A., Libeskind N.~I., Knollmann S.~R., Yepes G., Gottl{\"o}ber S., Hoffman Y., 2010, MNRAS, 405, 1119 
\bibitem[\protect\citeauthoryear{Knebe et al.}{2008b}]{2008MNRAS.388L..34K} Knebe A., Yahagi H., Kase H., Lewis G., Gibson B.~K., 2008b, MNRAS, 388, L34 
\bibitem[\protect\citeauthoryear{Kroupa et al.}{2010}]{2010A&A...523A..32K} Kroupa P., et al., 2010, A\&A, 523, A32 
\bibitem[\protect\citeauthoryear{Kroupa, Theis, \& Boily}{2005}]{2005A&A...431..517K} Kroupa P., Theis C., Boily C.~M., 2005, A\&A, 431, 517 
\bibitem[\protect\citeauthoryear{Kuhlen, Diemand, \& Madau}{2007}]{2007ApJ...671.1135K} Kuhlen M., Diemand J., Madau P., 2007, ApJ, 671, 1135 
\bibitem[\protect\citeauthoryear{Lambas et al.}{2012}]{2012A&A...539A..45L} Lambas D.~G., Alonso S., Mesa V., O'Mill A.~L., 2012, A\&A, 539, A45 
\bibitem[\protect\citeauthoryear{Lambas et al.}{2003}]{2003MNRAS.346.1189L} Lambas D.~G., Tissera P.~B., Alonso M.~S., Coldwell G., 2003, MNRAS, 346, 1189 
\bibitem[\protect\citeauthoryear{Lee, Kang, \& Jing}{2005}]{2005ApJ...629L...5L} Lee J., Kang X., Jing Y.~P., 2005, ApJ, 629, L5 
\bibitem[\protect\citeauthoryear{Lee et al.}{2008}]{2008MNRAS.389.1266L} Lee J., Springel V., Pen U.-L., Lemson G., 2008, MNRAS, 389, 1266 
\bibitem[\protect\citeauthoryear{Li et al.}{2013}]{2013ApJ...770L..12L} Li C., Jing Y.~P., Faltenbacher A., Wang J., 2013, ApJ, 770, L12 
\bibitem[\protect\citeauthoryear{Libeskind et al.}{2007}]{2007MNRAS.374...16L} Libeskind N.~I., Cole S., Frenk C.~S., Okamoto T., Jenkins A., 2007, MNRAS, 374, 16 
\bibitem[\protect\citeauthoryear{Libeskind et al.}{2009}]{2009MNRAS.399..550L} Libeskind N.~I., Frenk C.~S., Cole S., Jenkins A., Helly J.~C., 2009, MNRAS, 399, 550 
\bibitem[\protect\citeauthoryear{Libeskind et al.}{2016}]{2016ApJ...830..121L} Libeskind N.~I., Guo Q., Tempel E., Ibata R., 2016, ApJ, 830, 121 
\bibitem[\protect\citeauthoryear{Libeskind et al.}{2015}]{2015MNRAS.452.1052L} Libeskind N.~I., Hoffman Y., Tully R.~B., Courtois H.~M., Pomar{\`e}de D., Gottl{\"o}ber S., Steinmetz M., 2015, MNRAS, 452, 1052 
\bibitem[\protect\citeauthoryear{Libeskind et al.}{2014}]{2014MNRAS.443.1274L} Libeskind N.~I., Knebe A., Hoffman Y., Gottl{\"o}ber S., 2014, MNRAS, 443, 1274 
\bibitem[\protect\citeauthoryear{Libeskind et al.}{2011}]{2011MNRAS.411.1525L} Libeskind N.~I., Knebe A., Hoffman Y., Gottl{\"o}ber S., Yepes G., Steinmetz M., 2011, MNRAS, 411, 1525 
\bibitem[\protect\citeauthoryear{Libeskind et al.}{2018}]{2018MNRAS.473.1195L} Libeskind N.~I., et al., 2018, MNRAS, 473, 1195 
\bibitem[\protect\citeauthoryear{Limousin et al.}{2013}]{2013SSRv..177..155L} Limousin M., Morandi A., Sereno M., Meneghetti M., Ettori S., Bartelmann M., Verdugo T., 2013, SSRv, 177, 155 
\bibitem[\protect\citeauthoryear{Liu et al.}{2011}]{2011ApJ...733...62L} Liu L., Gerke B.~F., Wechsler R.~H., Behroozi P.~S., Busha M.~T., 2011, ApJ, 733, 62 
\bibitem[\protect\citeauthoryear{Lynden-Bell}{1976}]{1976MNRAS.174..695L} Lynden-Bell D., 1976, MNRAS, 174, 695 
\bibitem[\protect\citeauthoryear{M{\"u}ller et al.}{2018}]{2018Sci...359..534M} M{\"u}ller O., Pawlowski M.~S., Jerjen H., Lelli F., 2018, Sci, 359, 534 
\bibitem[\protect\citeauthoryear{McKay et al.}{2002}]{2002ApJ...571L..85M} McKay T.~A., et al., 2002, ApJ, 571, L85 
\bibitem[\protect\citeauthoryear{Mesa et al.}{2014}]{2014MNRAS.438.1784M} Mesa V., Duplancic F., Alonso S., Coldwell G., Lambas D.~G., 2014, MNRAS, 438, 1784 
\bibitem[\protect\citeauthoryear{Michel-Dansac et al.}{2008}]{2008MNRAS.386L..82M} Michel-Dansac L., Lambas D.~G., Alonso M.~S., Tissera P., 2008, MNRAS, 386, L82 
\bibitem[\protect\citeauthoryear{O'Mill et al.}{2012}]{2012MNRAS.421.1897O} O'Mill A.~L., Duplancic F., Garc{\'{\i}}a Lambas D., Valotto C., Sodr{\'e} L., 2012, MNRAS, 421, 1897 

\bibitem[\protect\citeauthoryear{Pajowska et al.}{2018}]{2018arXiv180802573P} Pajowska P., Godlowski W., Zhu Z.-H., Popiela J., Panko E., Flin P., 2018, arXiv, arXiv:1808.02573 

\bibitem[\protect\citeauthoryear{Patton et al.}{2011}]{2011MNRAS.412..591P} Patton D.~R., Ellison S.~L., Simard L., McConnachie A.~W., Mendel J.~T., 2011, MNRAS, 412, 591 
\bibitem[\protect\citeauthoryear{Patton et al.}{2016}]{2016MNRAS.461.2589P} Patton D.~R., Qamar F.~D., Ellison S.~L., Bluck A.~F.~L., Simard L., Mendel J.~T., Moreno J., Torrey P., 2016, MNRAS, 461, 2589 
\bibitem[\protect\citeauthoryear{Patton et al.}{2013}]{2013MNRAS.433L..59P} Patton D.~R., Torrey P., Ellison S.~L., Mendel J.~T., Scudder J.~M., 2013, MNRAS, 433, L59 
\bibitem[\protect\citeauthoryear{Pawlowski, Pflamm-Altenburg, \& Kroupa}{2012}]{2012MNRAS.423.1109P} Pawlowski M.~S., Pflamm-Altenburg J., Kroupa P., 2012, MNRAS, 423, 1109 
\bibitem[\protect\citeauthoryear{Pawlowski, Ibata, \& Bullock}{2017}]{2017ApJ...850..132P} Pawlowski M.~S., Ibata R.~A., Bullock J.~S., 2017, ApJ, 850, 132 
\bibitem[\protect\citeauthoryear{Pawlowski \& Kroupa}{2014}]{2014ApJ...790...74P} Pawlowski M.~S., Kroupa P., 2014, ApJ, 790, 74 
\bibitem[\protect\citeauthoryear{Pawlowski, Kroupa, \& Jerjen}{2013}]{2013MNRAS.435.1928P} Pawlowski M.~S., Kroupa P., Jerjen H., 2013, MNRAS, 435, 1928 
\bibitem[\protect\citeauthoryear{Pereira \& Kuhn}{2005}]{2005ApJ...627L..21P} Pereira M.~J., Kuhn J.~R., 2005, ApJ, 627, L21 
\bibitem[\protect\citeauthoryear{Pereira, Bryan, \& Gill}{2008}]{2008ApJ...672..825P} Pereira M.~J., Bryan G.~L., Gill S.~P.~D., 2008, ApJ, 672, 825 
\bibitem[\protect\citeauthoryear{Perez et al.}{2009}]{2009MNRAS.399.1157P} Perez J., Tissera P., Padilla N., Alonso M.~S., Lambas D.~G., 2009, MNRAS, 399, 1157 
\bibitem[\protect\citeauthoryear{Prada et al.}{2003}]{2003ApJ...598..260P} Prada F., et al., 2003, ApJ, 598, 260 
\bibitem[\protect\citeauthoryear{Sanders \& Evans}{2017}]{2017MNRAS.472.2670S} Sanders J.~L., Evans N.~W., 2017, MNRAS, 472, 2670 
\bibitem[\protect\citeauthoryear{Satyapal et al.}{2014}]{2014MNRAS.441.1297S} Satyapal S., Ellison S.~L., McAlpine W., Hickox R.~C., Patton D.~R., Mendel J.~T., 2014, MNRAS, 441, 1297 
\bibitem[\protect\citeauthoryear{Schneider et al.}{2013}]{2013MNRAS.433.2727S} Schneider M.~D., et al., 2013, MNRAS, 433, 2727 
\bibitem[\protect\citeauthoryear{Schneider \& Bridle}{2010}]{2010MNRAS.402.2127S} Schneider M.~D., Bridle S., 2010, MNRAS, 402, 2127 
\bibitem[\protect\citeauthoryear{Scudder et al.}{2015}]{2015MNRAS.449.3719S} Scudder J.~M., Ellison S.~L., Momjian E., Rosenberg J.~L., Torrey P., Patton D.~R., Fertig D., Mendel J.~T., 2015, MNRAS, 449, 3719 
\bibitem[\protect\citeauthoryear{Scudder et al.}{2012}]{2012MNRAS.426..549S} Scudder J.~M., Ellison S.~L., Torrey P., Patton D.~R., Mendel J.~T., 2012, MNRAS, 426, 549 
\bibitem[\protect\citeauthoryear{Shi, Wang, \& Mo}{2015}]{2015ApJ...807...37S} Shi J., Wang H., Mo H.~J., 2015, ApJ, 807, 37 
\bibitem[\protect\citeauthoryear{Shimasaku et al.}{2001}]{2001AJ....122.1238S} Shimasaku K., et al., 2001, AJ, 122, 1238 
\bibitem[\protect\citeauthoryear{Sif{\'o}n et al.}{2015}]{2015A&A...575A..48S} Sif{\'o}n C., Hoekstra H., Cacciato M., Viola M., K{\"o}hlinger F., van der Burg R.~F.~J., Sand D.~J., Graham M.~L., 2015, A\&A, 575, A48 
\bibitem[\protect\citeauthoryear{Singh, Mandelbaum, \& More}{2015}]{2015MNRAS.450.2195S} Singh S., Mandelbaum R., More S., 2015, MNRAS, 450, 2195 
\bibitem[\protect\citeauthoryear{Smith et al.}{2016}]{2016ApJ...818...11S} Smith R., Duc P.~A., Bournaud F., Yi S.~K., 2016, ApJ, 818, 11 
\bibitem[\protect\citeauthoryear{Tasca \& White}{2011}]{2011A&A...530A.106T} Tasca L.~A.~M., White S.~D.~M., 2011, A\&A, 530, A106 
\bibitem[\protect\citeauthoryear{Teklu et al.}{2017}]{2017MNRAS.472.4769T} Teklu A.~F., Remus R.-S., Dolag K., Burkert A., 2017, MNRAS, 472, 4769 
\bibitem[\protect\citeauthoryear{Tempel et al.}{2015}]{2015MNRAS.450.2727T} Tempel E., Guo Q., Kipper R., Libeskind N.~I., 2015, MNRAS, 450, 2727 
\bibitem[\protect\citeauthoryear{Tempel et al.}{2014}]{2014MNRAS.437L..11T} Tempel E., Libeskind N.~I., Hoffman Y., Liivam{\"a}gi L.~J., Tamm A., 2014, MNRAS, 437, L11 
\bibitem[\protect\citeauthoryear{Tempel et al.}{2016}]{2016A&C....16...17T} Tempel E., Stoica R.~S., Kipper R., Saar E., 2016, A\&C, 16, 17 
\bibitem[\protect\citeauthoryear{Tempel et al.}{2014}]{2014MNRAS.438.3465T} Tempel E., Stoica R.~S., Mart{\'{\i}}nez V.~J., Liivam{\"a}gi L.~J., Castellan G., Saar E., 2014, MNRAS, 438, 3465 
\bibitem[\protect\citeauthoryear{Tempel \& Tamm}{2015}]{2015A&A...576L...5T} Tempel E., Tamm A., 2015, A\&A, 576, L5 
\bibitem[\protect\citeauthoryear{Tempel et al.}{2014}]{2014A&A...566A...1T} Tempel E., et al., 2014, A\&A, 566, A1 
\bibitem[\protect\citeauthoryear{Tenneti et al.}{2014}]{2014MNRAS.441..470T} Tenneti A., Mandelbaum R., Di Matteo T., Feng Y., Khandai N., 2014, MNRAS, 441, 470 
\bibitem[\protect\citeauthoryear{Tenneti et al.}{2015}]{2015MNRAS.448.3522T} Tenneti A., Singh S., Mandelbaum R., di Matteo T., Feng Y., Khandai N., 2015, MNRAS, 448, 3522 
\bibitem[\protect\citeauthoryear{Thompson}{1976}]{1976ApJ...209...22T} Thompson L.~A., 1976, ApJ, 209, 22 
\bibitem[\protect\citeauthoryear{Tormen}{1997}]{1997MNRAS.290..411T} Tormen G., 1997, MNRAS, 290, 411 
\bibitem[\protect\citeauthoryear{van den Bosch et al.}{2004}]{2004MNRAS.352.1302V} van den Bosch F.~C., Norberg P., Mo H.~J., Yang X., 2004, MNRAS, 352, 1302 
\bibitem[\protect\citeauthoryear{Violino et al.}{2018}]{2018MNRAS.476.2591V} Violino G., Ellison S.~L., Sargent M., Coppin K.~E.~K., Scudder J.~M., Mendel T.~J., Saintonge A., 2018, MNRAS, 476, 2591 
\bibitem[\protect\citeauthoryear{Vitvitska et al.}{2002}]{2002ApJ...581..799V} Vitvitska M., Klypin A.~A., Kravtsov A.~V., Wechsler R.~H., Primack J.~R., Bullock J.~S., 2002, ApJ, 581, 799 
\bibitem[\protect\citeauthoryear{Wang et al.}{2005}]{2005MNRAS.364..424W} Wang H.~Y., Jing Y.~P., Mao S., Kang X., 2005, MNRAS, 364, 424 
\bibitem[\protect\citeauthoryear{Wang et al.}{2018}]{2018ApJ...866..138W} Wang P., Guo Q., Kang X., Libeskind N.~I., 2018, ApJ, 866, 138
\bibitem[\protect\citeauthoryear{Wang \& Kang}{2018}]{2018MNRAS.473.1562W} Wang P., Kang X., 2018, MNRAS, 473, 1562 
\bibitem[\protect\citeauthoryear{Wang \& Kang}{2017}]{2017MNRAS.468L.123W} Wang P., Kang X., 2017, MNRAS, 468, L123 
\bibitem[\protect\citeauthoryear{Wang et al.}{2018}]{2018ApJ...859..115W} Wang P., Luo Y., Kang X., Libeskind N.~I., Wang L., Zhang Y., Tempel E., Guo Q., 2018, ApJ, 859, 115 
\bibitem[\protect\citeauthoryear{Wang et al.}{2014}]{2014ApJ...786....8W} Wang Y.~O., Lin W.~P., Kang X., Dutton A., Yu Y., Macci{\`o} A.~V., 2014, ApJ, 786, 8 
\bibitem[\protect\citeauthoryear{Wei et al.}{2018}]{2018ApJ...853...25W} Wei C., et al., 2018, ApJ, 853, 25 
\bibitem[\protect\citeauthoryear{Yang et al.}{2006}]{2006MNRAS.369.1293Y} Yang X., van den Bosch F.~C., Mo H.~J., Mao S., Kang X., Weinmann S.~M., Guo Y., Jing Y.~P., 2006, MNRAS, 369, 1293 
\bibitem[\protect\citeauthoryear{Yang et al.}{2014}]{2014JGRG..119.2245Y} Yang Y., Long D., Guan H., Scanlon B.~R., Simmons C.~T., Jiang L., Xu X., 2014, JGRG, 119, 2245 
\bibitem[\protect\citeauthoryear{York et al.}{2000}]{2000AJ....120.1579Y} York D.~G., et al., 2000, AJ, 120, 1579 
\bibitem[\protect\citeauthoryear{Zel'dovich}{1970}]{1970A&A.....5...84Z} Zel'dovich Y.~B., 1970, A\&A, 5, 84 
\bibitem[\protect\citeauthoryear{Zentner et al.}{2005}]{2005ApJ...629..219Z} Zentner A.~R., Kravtsov A.~V., Gnedin O.~Y., Klypin A.~A., 2005, ApJ, 629, 219 





\end{thebibliography}
\end{document}